\documentclass{JHEP3}
\usepackage{graphicx}

\def\bea{\begin{eqnarray}}
\def\eea{\end{eqnarray}}

\newcommand{\mtau}{m_\tau}

\newcommand{\Sew}{{S_{\rm EW}}}

\newcommand{\Kbar    }{\kern 0.2em\overline{\kern -0.2em K}{}}

\preprint{LPT-ORSAY 10-10\\ LAL 10-17}
\title{A note on renormalon models\\ for the determination of $\alpha_s(M_\tau)$}
\author{S. Descotes-Genon\\ Laboratoire de Physique Th\'eorique,\\
CNRS \& Univ. Paris-Sud 11 (UMR 8627), 91405 Orsay Cedex, France}
\author{B. Malaescu\\Laboratoire de l'Acc\'el\'erateur Lin\'eaire,\\ IN2P3/CNRS \& Univ. Paris-Sud 11 (UMR 8607),91405, Orsay Cedex, France}
\abstract{The tau hadronic width provides a determination of the strong coupling constant $\alpha_s$ at low energies, since it can be related to a weighted integral of the Adler function in the complex energy plane. Using Operator Product Expansion, one sees that the sensitivity to $\alpha_s$ comes from the perturbative contribution, which can be obtained by integrating the perturbative expansion of the Adler function. Two different prescriptions proposed to perform this integral, called Fixed-Order Perturbation Theory and Contour-Improved Perturbation Theory (FOPT and CIPT), yield  different results for the strong coupling constant. Recently, models for the Adler function based on renormalon calculus have been proposed to determine which of the two methods is the most accurate, by comparing the resulting asymptotic series with the true value of the integral. We discuss the assumptions of such ansatz and the determination of their free parameters. We show that variations of this renormalon ansatz can yield opposite conclusions concerning the comparison of CIPT versus FOPT, and that such models
are not constrained enough to provide a definite answer on this issue or to be exploited for
a high-precision determination of $\alpha_s(m_\tau^2)$.
}

\begin{document}

\section{Introduction}

A precise assessment of the fundamental parameters of the Standard Model is mandatory to test 
the consistency of the theory and to exploit its full predictive potential, with the aim of identifying discrepancies indicating loopholes in our understanding of its dynamics or providing signs of New Physics. The coupling constant $\alpha_s$ is the central ingredient of the strong sector of the theory, which can be determined at different scales using a large range of processes~\cite{Bethke:2009jm}. At low energies, a precise value of $\alpha_S(M_\tau^2)$ can be extracted from the $\tau$ hadronic width $R_\tau$
\begin{equation}
R_\tau=\frac{\Gamma[\tau\to {\rm hadrons}\ \nu_\tau (\gamma)]}{\Gamma[\tau\to e\bar\nu_e\nu_\tau (\gamma)}
\end{equation}
as well as its moments~\cite{Narison:1988ni,Braaten:1988ea}, which can be determined from the analysis performed by the LEP experiments~\cite{Davier:2005xq}, and complemented with further experimental input from $B$-factories.

The theoretical description of this quantity can be obtained by relating this decay width with
an integral of the $\tau$ spectral functions, corresponding to the imaginary part of two-point correlators
of hadronic currents. Exploiting the analytic properties of theses correlators, one can reexpress the integral as a contour integral of the Adler function $D$ along a circle of radius $|s|=M_\tau^2$ in the complex plane of center-of-mass energy. One can then use the Operator Product Expansion of the correlator~\cite{Shifman:1978bx} to compute the various contributions to the decay width~\cite{Narison:1988ni,Braaten:1988ea,Braaten:1991qm,Le Diberder:1992te,Le Diberder:1992fr}:
\begin{equation}
R_{\tau,V/A}=\frac{N_c}{2} S_{EW} |V_{ud}|^2 [1+\delta^{(0)}+\delta'_{EW}+
  \sum_{D\geq 2} \delta^{(D)}_{ud,V/A}]
\end{equation}
with $S_{EW}$ collects short-distance electroweak corrections~\cite{Marciano:1988vm,Braaten:1990ef}, $\delta^{(0)}$ stems from the purely perturbative part of the expansion, defined in the chiral limit, while mass corrections are incorporated in (higher) $D$-dimensional contributions $\delta^{(D)}_{ud,V/A}]$ containing also the contributions from the condensates.

The main sensitivity to $\alpha_s$ comes from $\delta^{(0)}$, defined as
\begin{equation}\label{eq:delta0}
1+\delta^{(0)}=-2 \pi i \oint_{|s|=s_0} \frac{ds}{s}w(s) D(s), \qquad w(s)=1-2\frac{s}{s_0}+2\left(\frac{s}{s_0}\right)^3-\left(\frac{s}{s_0}\right)^4
\end{equation}
Since we know only the first orders of the perturbative expansion of the Adler function for a given (real) value of the scale, the method used to compute the contour integral turns out to be important, in order to control (and suppress) the size of unknown higher orders. In particular, two different rules have been proposed, called Fixed-Order Perturbation Theory (FOPT) and Contour-Improved Perturbation Theory (CIPT)~\cite{Braaten:1991qm,Le Diberder:1992te,Le Diberder:1992fr}. We will recall below the most salient features of the two approaches. It turns out that
the increasing experimental accuracy and the determination of the $O(\alpha_s^4)$ term in the perturbative series of the Adler function has proved the two different approaches to differ significantly and to induce a significant systematic uncertainty in the determination of the strong coupling constant.

Several studies have been performed to determine which of the two methods (if any) is to be
preferred. A first ("internal") way of dealing with this issue consists in comparing the two methods to determine if one has a more regular and stable behaviour than the other, proving thus its robustness.
For instance, in ref.~\cite{Davier:2008sk}, our current knowledge of the perturbative series for the Adler function led us to conclude that CIPT showed a much better stability than FOPT as far as the dependence of the scale defining the integration contour is concerned, and that the FOPT integrand showed a pathological behaviour once one gets close to the end of the integration circle $s=M_\tau^2$. In ref.~\cite{Menke:2009vg}, a similar study was performed, where the reference point for the Taylor expansion of the coupling constant used in FOPT was varied on a circle of radius $M_\tau$ in the complex energy plane, hinting at a strong dependence of the FOPT value due to a large impact of logarithmic corrections in the Taylor expansion. However, let us stress that these conclusions rely on the determination of $\alpha_s$ at a reference point in the complex plane by iterating the RGE step by step along the contour of integration (in a similar way to the CIPT method), and thus use a "non-canonical" version of FOPT including elements of the CIPT philosophy. 
A model for higher orders in the perturbative expansion of the Adler function~\cite{Jamin:2005ip} suggested that FOPT could actually oscillate towards the (more stable) CIPT value once higher-order are included.

One can also opt for a different ("external") approach, where the true value of the Adler function and its perturbative expansion are assumed to be known. One can then determine if the true value of the integral is approached by one of the two methods when one starts increasing the accuracy of the perturbative expansion (before the series becomes asymptotic). A particular model was proposed more recently in ref.~\cite{Beneke:2008ad}, based on renormalon calculus~\cite{Ball:1995ni,Neubert:1995gd,Beneke:1998ui,Beneke:2000kc}. The first observation consists in the fact that perturbative series like
\begin{equation}
D(Q^2)=1+\sum_{n=1}^\infty c_{n,1} a_Q^n \qquad a_Q=\alpha_s(Q)/\pi
\end{equation}
are at best asymptotic ones, with a zero convergence radius in $a_Q$, but their Borel transform can be defined as
\begin{equation}\label{eq:borelseries}
B[D](t)=\sum_{n=0}^\infty c_{n+1,1} \frac{t^n}{\pi^{n+1} n!}
\end{equation}
with improved convergence properties, and in particular a non-vanishing convergence radius. If $B[D](t)$ has no singularity for t real and positive, and does not increase too quickly at infinity, one can define the Borel sum:
\begin{equation}\label{eq:borelsum}
\tilde{D}(a)=\int_0^\infty dt\ e^{t/a} B[D](t)
\end{equation}
The $\tilde{D}(a)$ has the same perturbative expansion in powers of $a$ as $D(Q^2)$ in $\alpha_s(Q)$.
Actually, the Borel transform is expected to have singularities along the real axis, for borh 
positive and negative values of $t$. The former are called infrared renormalons, and are related to power corrections and condensates in the Operator Product Expansion of the Adler function, whereas the latter are called ultraviolet renormalons and are related to the large-order (and often oscillatory) behaviour of the series at higher orders. The presence of infrared renormalons requires to give a prescription to avoid the singularities in eq.\ref{eq:borelsum}) (most often the principal value).

In ref.~\cite{Beneke:2008ad}, a particular ansatz $B[D]$ was proposed to describe the first singularities close to the edge of the domain of convergence as a sum of "poles" with fractional powers (actually cuts). The free parameters were determined from the first orders of the perturbative expansion and the known properties of the Operator Product Expansion. This ansatz was used to compute $\tilde{D}$ at arbitrary orders in perturbation theory, and then the integral $\delta^{(0)}$ using either FOPT or CIPT. In this case, when one increases the order, the evolution of the FOPT value of $\delta^{(0)}$ exhibits a plateau in agreement with the value obtained from the Borel sum $\tilde{D}$.

This analysis is based on several assumptions. 
Because $D$ and $\tilde{D}$ share the same perturbative expansion for real positive values of the coupling constant, the latter is expected to yield the "true" value of the Adler function for arbitrary (complex) values of the strong coupling constant in its convergence radius. 
One could then determine the "true" value of $\delta^{0}$ by integrating the Borel sum $D(\alpha(s))$ over a circle in the complex $s$-plane, i.e. for complex values of $\alpha_s$.
The ansatz for $B[D]$ contains the three singularities that are relevant not only at high orders (where they are dominated by the first ultraviolet renormalons), but also at intermediate orders (where they are dominated by the first infrared renormalons), and even at low orders (since the first five orders of the perturbative expansion of $D$ are used to determine the free parameters in $B[D]$).

In refs.~\cite{Caprini:1998wg,Caprini:2009vf,Caprini:2009nr}, some aspects of this ansatz were discussed to map the Borel parameter $t$ into another parameter $w$ so that the cut plane along the real axis would be mapped into a disc of unit radius. It was argued that the series eq.~(\ref{eq:borelseries}) has optimal convergence properties once expressed in $w$, which would select this variable as the appropriate one to discuss renormalon models. Even though the main ingredients are the same (structure of the singularities, first orders of the perturbative series), the choice of $w$ rather than $t$ modifies the (non-singular) structure of the series eq.~(\ref{eq:borelseries}) and it was enough to alter significantly the outcome of the analysis.

The present note aims at investigating other aspects of this ansatz. In Sec.~2, we recall the theoretical framework leading from the $\tau$ decay width to the perturbative contribution
$\delta^{(0)}$. In Sec.~3, we present the ansatz used in ref.~\cite{Beneke:2008ad}, describing how the free parameters of the model are fixed. In Sec.~4, we discuss how the uncertainty on the Borel sum $\tilde{D}$ can be underestimated when one use it to compute $\delta^{0}$, depending on the way the singularities related to infrared renormalons are avoided. In Sec.~5, we discuss in some detail the FOPT and CIPT prescriptions, and we identify two different ways of applying FOPT, leading to a further uncertainty for this prescription. In Sec.~6, we discuss how the free parameters of the model are fixed, based on a high-order expansion of the renormalon contributions which is affected by potentially significant corrections. We mimic the effect of
such corrections by introducing a quadratic term in the series defining the Borel series, and study the impact on the FOPT/CIPT discussion. In Sec.~7, we study the notion of pole dominance, which was used in ref.~\cite{Beneke:2008ad} to favour their ansatz, and we see that several definitions could be imagined for such a dominance. In Sec.~8, we discuss the role of anomalous dimensions in the
results of the discussion, in order to treat the first two infrared renormalons at the same level of detail. In Sec.~9, we show that the agreement of the FOPT or CIPT value with the Borel sum depends on the chosen weight, and is thus related to enhancement or cancellation of some 
parts of the integration contour. Finally, in Sec.~10, we summarise and conclude our study.

\section{Theoretical framework}

From the theoretical point of view, the $\tau$ decay width $R_\tau$ can be described in terms of its contributions from non-strange vector $ud$, non-strange axial $ud$ and strange $us$
\begin{equation}
R_\tau=R_{\tau,V}+R_{\tau,A}+R_{\tau,S}.
\end{equation}
One can relate each of these to the corresponding spectral functions through
\begin{equation}
R_\tau=12\pi S_{EW}\int_0^{M_\tau^2} \left(1-\frac{s}{M_\tau^2}\right)^2
  \left[\left(1+2\frac{s}{M_\tau^2}\right)Im\ \Pi^{(1)}(s)+Im\ \Pi^{(0)}(s)\right]
\end{equation}
and the spectral functions are related to the imaginary part of the correlators:
\begin{equation}
\Pi^{(J)}=|V_{ud}|^2[\Pi^{V(J)}_{ud}(s)+\Pi^{A(J)}_{ud}(s)]
         +|V_{us}|^2[\Pi^{V(J)}_{us}(s)+\Pi^{A(J)}_{us}(s)]
\end{equation}
combining two-point correlators of hadronic current with given angular momentum. They
are defined from the correlator defined in QCD:
\begin{equation}
\Pi^X_{\mu\nu,uD}(p)=i\int dx e^{ipx} \langle 0|TJ^X_{\mu,uD}(x)J^X_{\mu,uD}(0)^\dag|0\rangle
\end{equation}
where $X=V$ or $A$, $D=d$ or $s$, and the hadronic currents are 
$J^V_{\mu,uD}=\bar{D}\gamma_\mu u$ and $J^A_{\mu,uD}=\bar{D}\gamma_\mu\gamma_5 u$. This correlator
can be decomposed according to angular momentum:
\begin{equation}
\Pi^X_{\mu\nu,uD}(p)=(p_\mu p_\nu - g_{\mu\nu} p^2) \Pi^{X(1)}_{uD}(p^2)
  + p_\mu p_\nu \Pi^{X(0)}_{uD}(p^2)
\end{equation}

Since the correlator has only a singularity along the positive real axis,
one can deform the integration contour and reexpress this quantity as
\begin{eqnarray}\label{eq:continteg}
R_\tau &=& 6 i \pi S_{EW} \oint_{|s|=M_\tau^2} \frac{ds}{M_\tau^2} 
  \left(1-\frac{s}{M_\tau^2}\right)^2
  \left[\left(1+2\frac{s}{M_\tau^2}\right)\Pi^{(1)}(s)+\Pi^{(0)}(s)\right]\\
&=& -i\pi S_{EW}\oint_{|x|=1} \frac{dx}{x} (1-x)^3 \left[3(1+x)D^{(1+0)}(M_\tau^2 x)+4 D^{(0)}(M_\tau^2x)\right]
\end{eqnarray} 
where we have rewritten the integral in terms of the Adler functions
\begin{equation}
D^{(1+0)}=-s\frac{d}{ds}[\Pi^{(1+0}(s)] \qquad D^{(0)}=
  \frac{s}{M_\tau^2}\frac{d}{ds}[s\Pi^{(0)}(s)]
\end{equation}

If one performs the operator product expansion of $D$, one can write down
an expression of $R_\tau$ of the form
\begin{equation}
R_{\tau,V/A}=\frac{N_c}{2} S_{EW} |V_{ud}|^2 [1+\delta^{(0)}+\delta'_{EW}+
  \sum_{D\geq 2} \delta^{(D)}_{ud,V/A}]
\end{equation}
In the chiral limit considered to compute the perturbative contribution, $D^{(0)}=0$ and one can consider only $D^{(1+0)}$ which is independent of the renormalisation scale and has an expansion of the type:
\begin{equation} \label{eq:pertAdler}
D_V^{(1+0)}(Q^2)=\frac{N_c}{12\pi^2}\sum_{n=0}^\infty c_{n,1} a_Q^n
\end{equation}
with $a_Q=\alpha_s(Q)/\pi$. The first few orders are known:
\begin{equation}
c_{0,1}=c_{1,1}=1 \qquad c_{2,1}=1.640 \qquad c_{3,1}=6.371 \qquad c_{4,1}=49.076
\end{equation}
The value of $c_{5,1}$ is still unknown, and a very frequent assumption consists in taking a geometric progression for these numbers, leading to $c_{5,1}\simeq 283$.

The issue now consists in determining how to perform the integration over the contour for the perturbative contribution in the most accurate way. This issue is of importance here for two different reasons:
\begin{itemize}
\item The main outcome of the analysis consists in the determination of $\alpha_s(m_\tau^2)$, which is involved mainly through the perturbative contribution to the computation of the $\tau$-width
\item The Operator Product Expansion is expected to break down once one gets closer to the cut corresponding to the physical region, and thus its integration around a circular contour in the complex plane should be considered with a particular care
\end{itemize}

\section{Model of renormalons} \label{sec:renormmodel}

In ref.~\cite{Beneke:2008ad}, a renormalon model was presented to describe higher orders for the Adler function $\hat{D}$ eq.~(\ref{eq:pertAdler}), and thus to compare different integration methods once higher order are taken into account. The ansatz
consisted in one ultraviolet renormalon, corresponding to the sign-alternating divergence expected at higher orders in the perturbative series of $D$, and two infrared renormalons, mirroring the presence of condensates of dimension 4 (gluon condensate) and 6 (higher order quark and gluon condensates) in the OPE of the Adler function taken in the chiral limit~\cite{Ball:1995ni,Neubert:1995gd,Beneke:1998ui,Beneke:2000kc}.

In the Borel plane, the ansatz for the Adler function has the following representation
\begin{equation}\label{eq:ansatzBJ}
B[\hat{D}](u)=B[\hat{D}_1^{UV}](u)+B[\hat{D}_2^{IR}](u)+B[\hat{D}_3^{IR}](u)
      +d_0^{PO}+d_1^{PO} u
\end{equation}
where each renormalon "pole" has a cut singularity of the form
\begin{equation}\label{eq:cutpole}
B[\hat{D}_p^{X}](u)=\frac{d_p^X}{(p\mp u)^{1+\gamma}}[1+b_1(p\mp u)+b_2(p\mp u)^2+\ldots]
\end{equation}
where the negative sign corresponds to an infrared (IR) renormalon, and the positive one to an ultraviolet (UV) renormalon.

The corresponding model for the Adler function is given by 
\begin{equation}\label{eq:resumborel}
\hat{D}(\alpha)=\int_0^\infty dt e^{-t/\alpha} B[\hat{D}](t(u))
\end{equation}
with $t=\pi u/\beta_0$. The real part of this integral is expected to yield the "true" value of the perturbative series, whereas its imaginary part (divided by $\pi$) should provide an estimate of the uncertainty, attached to the way one treats the singularities related to IR renormalons.

The corresponding perturbative series can be worked out as outlined in sec.5 of ref.~\cite{Beneke:2008ad}
\begin{eqnarray}\label{eq:perturbexp}
\hat{D}_p^{X}(a_Q)&=&\frac{\pi d_p^{X}}{p^{1+\gamma} \Gamma(1+\gamma)} 
  \sum_{n=0}^\infty \Gamma(n+1+\gamma) \left(\pm \frac{\beta_0}{p}\right)^n a_Q^{n+1}\\
 && \qquad \times \left[1+\frac{p}{\beta_0} \frac{(b_1+c_1)}{n+\gamma}+\left(\frac{p}{\beta_0}\right)^2 \frac{(b_2+b_1c_1+ c_2)}{(n+\gamma)(n+\gamma-1)}+O\left(\frac{1}{n^3}\right)\right]
 \nonumber
\end{eqnarray}
where $a_Q=\alpha_s(Q)/\pi$. The values of the coefficients can be determined in the case of IR renormalons because their structure is connected with the contributions of the non-perturbative condensates occurring in the Operator Product Expansion of the correlator: $p$ is related to the naive dimension of the condensate, $\gamma$ to its anomalous dimension, the $b$ coefficients to the running of the strong coupling constant and the $c$ coefficients to the perturbative series multiplying the condensate in the OPE. An analytic continuation can then be performed to assume a similar connection between UV renormalons and higher-dimension operators.

The relative weight of the three renormalon contributions (indicated by $d_1^{UV},d_2^{IR},d_3^{IR}$) was fixed using the 3rd, 4th and 5th orders  in the expansion of the Adler function ($c_{3,1},c_{4,1},c_{5,1}$, the latter being set assuming a geometrical growth of the coefficients). All the coefficients $c$ in eq.~(\ref{eq:perturbexp}), related to the Wilson coefficients in the OPE, were set to zero apart from the coefficient $c_1$ for the $d_2^{IR}$ pole. The first two terms ($c_{1,1},c_{2,1}$) were not considered and are reproduced by adding an ad hoc term in the model (constant and linear terms in $u$).

\section{Uncertainty on the Borel integral}

One can use eq.~(\ref{eq:resumborel}) in order to derive a "resummed value" of the perturbative expansion. The presence of IR poles on the positive real axis means that we have to give a prescription for the integral, depending on whether we take the contour integral above or below the real axis, which yields:
\begin{equation}
\hat{D}(s)=\hat{D}_{PV}(s) \pm i \hat{D}_{pole}(s)
\end{equation}
corresponding to the principal value and the pole contribution of the integral in eq.~(\ref{eq:resumborel}). When $s$ is along the real axis, both $\hat{D}_{PV}$ and $\hat{D}_{pole}$ are real, but this does not remain the case in the complex plane. 

We perform the contour integral in the complex plane eq.~(\ref{eq:continteg}) in order to compute $\delta^{(0)}$:
\begin{equation}
\delta^{(0)}=\oint_{|s|=M_\tau^2} ds\ K(s)\ \hat{D}(s)=
\oint_{|s|=M_\tau^2} ds\ K(s)\ [\hat{D}_{PV}(s) \pm i \hat{D}_{pole}(s)]
\end{equation}
where $Re\ \hat{D}_{PV}(s_0\exp[i\phi])$, $Re\ \hat{D}_{pole}(s_0\exp[i\phi])$ and
$Re\ K(s_0\exp[i\phi])$ are even functions of $\phi$, whereas the imaginary parts are odd functions of the same variable. Therefore the value of $\delta^{(0)}$ is obtained from
\begin{equation}
\delta^{(0)}=\oint_{|s|=M_\tau^2}\!\!\!\! ds\ K(s)\ \hat{D}_{PV}(s)=
 \oint_{|s|=M_\tau^2}\!\!\!\! ds\ [Re\ K\ Re\ \hat{D}_{PV}-Im\ K\ Im\ \hat{D}_{PV}](s)
\end{equation}
The imaginary part of the integral evaluated with the principal value prescription (and divided by $\pi$) is sometimes taken as an estimate of the uncertainty on the value of the Borel integral~\cite{Beneke:2008ad}:
\begin{equation}
Err\ \delta^{(0)}=\frac{1}{\pi}\oint_{|s|=M_\tau^2}\!\!\!\! ds\ K(s)\ \hat{D}_{pole}(s)
 =\oint_{|s|=M_\tau^2}\!\!\!\! ds\ [Re\ K\ Re\ \hat{D}_{pole}-Im\ K\ Im\ \hat{D}_{pole}](s)
\end{equation}

Once several IR poles are included in the model, one must decide how to combine the contributions 
from $\hat{D}_{pole}^{IR2}$ and $\hat{D}_{pole}^{IR3}$ in $Err\ \delta^{(0)}$. Ref.~\cite{Beneke:2008ad} takes the sum of the two contributions with a relative sign maximising the error, i.e.
\begin{equation}
Err\ \delta^{(0)}=\frac{1}{\pi}\left|\oint_{|s|=M_\tau^2} ds\ K(s)\ 
      [\hat{D}_{pole}^{IR2}(s)\pm\hat{D}_{pole}^{IR3}(s)]\right|
\end{equation}
This amounts to assuming correlations for the variation of the uncertainty when one moves along the circle in the complex plane (the relative sign is assumed to be the same for any position along the circle). It seems more conservative to assume an absence of correlations, taking:
\begin{equation}
Err\ \delta^{(0)}=\frac{1}{\pi}\oint_{|s|=M_\tau^2} ds\ K(s)\ 
      [|\hat{D}_{pole}^{IR2}(s)|+|\hat{D}_{pole}^{IR3}(s)|]
\end{equation}
This prescription tends to increase the error bar on the Borel integral in a significant way.

\section{Different methods of treating perturbation theory}

In refs.~\cite{Beneke:2008ad,Davier:2005xq,Le Diberder:1992te,Le Diberder:1992fr,Davier:2008sk} were discussed two main prescriptions to compute the integral eq.~(\ref{eq:delta0}) in terms of the perturbative expansion of the Adler function eq.~(\ref{eq:pertAdler}), called Fixed-Order Perturbation Theory (FOPT) and Contour-Improved Perturbation Theory (CIPT). 
The methods were actually presented differently in refs.\cite{Davier:2005xq,Le Diberder:1992te,Le Diberder:1992fr} and ref.\cite{Beneke:2008ad}, and we will recall the salient elements of both presentations.

The starting point of refs.\cite{Le Diberder:1992te,Le Diberder:1992fr,Davier:2005xq} is the solution of
the RGE which reads -- as defined in~\cite{vanRitbergen:1997va,Larin:1997qq}:
\begin{equation}\label{eq:RGE}
-\frac{da_\mu}{d\log(\mu^2)}=\sum_{k=0}^\infty \beta_k  a_\mu^{k+2}
\end{equation}
The full expressions for an arbitrary number of quark flavours ($n_f$) 
   are: $\beta_0 = \frac{1}{4}\left(11 - \frac{2}{3}n_f\right)$,
	 $\beta_1 = \frac{1}{16}\left(102 - \frac{38}{3}n_f\right)$,
	 $\beta_2 = \frac{1}{64}\left(\frac{2857}{2} - \frac{5033}{18}n_f 
     + \frac{325}{54}n_f^2\right)$, and
	$$\beta_3 = \frac{1}{256}\left[
	       \frac{149753}{6} + 3564\,\zeta_3 -
	       \left(\frac{1078361}{162} + \frac{6508}{27}\,\zeta_3\right)n_f \right. \,+\,
          \left. \left(\frac{50065}{162} + \frac{6472}{81}\,\zeta_3\right)n_f^2 +
	       \frac{1093}{729}n_f^3\right]$$
	where $\zeta_{i=\{3,4,5\}}=\{1.2020569,\pi^4/90,1.0369278\}$ are the Riemann 
	$\zeta$-functions. 
The standard perturbative method to compute the contour integral consists then of expanding
all the quantities up to a given power of $a_s(s_0)$. The starting point is the 
solution of the renormalisation group equation (RGE) for $a_s(s)$, which is expanded 
in a Taylor series of $\eta\equiv\ln(s/s_0)$ around the reference scale $s_0$~\cite{Davier:2005xq}
\begin{eqnarray}
\label{eq:astaylor}
&& a_s(s) =
  		a_s - 
		\beta_0 \eta a_s^2 + 
		\left(-\beta_1\eta + \beta_0^2 \eta^2\right) a_s^3 +
    	\bigg(-\beta_2\eta  
        + \frac{5}{2} \beta_0\beta_1 \eta^2 - \beta_0^3 \eta^3 \bigg) a_s^4 \nonumber \\ 
      &&   +\; \left(-\beta_3\eta  + \frac{3}{2} \beta_1^2\eta^2 + 3 \beta_0 \beta_2\eta^2  
		      - \frac{13}{3} \beta_0^2 \beta_1 \eta^3
                	+ \beta_0^4 \eta^4
				\right) a_s^5 \\ 
      && +\; \bigg( -\beta_4\eta + \frac{7}{2} \beta_1 \beta_2\eta^2 
		          + \frac{7}{2} \beta_0 \beta_3\eta^2  
						- \frac{35}{6} \beta_0 \beta_1^2 \eta^3 
               				- 6 \beta_0^2 \beta_2 \eta^3 
                           + \frac{77}{12} \beta_0^3 \beta_1 \eta^4
       			- \beta_0^5 \eta^5
				\bigg)a_s^6 
		+\mathcal{O}(\eta^6;a_s^7)~.  \nonumber 
\end{eqnarray}
Here the series has been reordered in powers of $a_s\equiv a_s(s_0)$ and 
we use the RGE $\beta$-function~(\ref{eq:RGE}). 
We insert
eq.(\ref{eq:astaylor}) up to  $a_s^5$ (since we know the $\beta$ coefficients only up to that order) into 
the integral eq.~(\ref{eq:delta0}), and we order the contributions according to their 
powers in $a_s$, so that we obtain the familiar expression for fixed-order perturbation 
theory (FOPT)~\cite{Le Diberder:1992te,Le Diberder:1992fr}
\begin{equation}
\label{eq:kngn}
   \delta^{(0)} = 
       \sum_{n=1}^\infty \left[\tilde{K}_n(\xi) + g_n(\xi)\right]
       a_s^n(\xi s_0)\,,
\end{equation}
where the $g_n$ are functions of $\tilde{K}_{m<n}$ and $\beta_{m<n-1}$, and of
elementary integrals with logarithms of power $m<n$ in the integrand. 

One can also perform the integral by performing a step-by-step integration along the circular contour, using the RGE solution 
eq.~(\ref{eq:astaylor}) to determine the value of the strong coupling constant at each point of integration. The initial point lies on the real axis (like for FOPT), and as one moves along the circular contour, the value of $\alpha_s$ at a given point is computed using RGE with the value obtained at the previous step. This second method was introduced in refs.~(\cite{Le Diberder:1992te,Le Diberder:1992fr}) and is called Contour-Improved Perturbation Theory (or CIPT). Its advantages were discussed in ref.~\cite{Davier:2008sk}.

In ref.~\cite{Beneke:2008ad}, a different angle was chosen to present the methods. The starting point consisted not in inserting the solution of the RGE (\ref{eq:astaylor}) into the perturbative expansion of the Adler function eq.~(\ref{eq:pertAdler}), but rather in exploiting the fact that the Adler function is independent of the renormalisation scale. Indeed if we start from 
the perturbative expression for $\Pi$:
\begin{equation}
\Pi_V^{(1+0)}(s)=-\frac{N_c}{12\pi^2}\sum_{n=0}^\infty a_\mu^n 
  \sum_{k=0}^{n+1} c_{n,k} \tilde\eta^k \qquad \tilde\eta=\log\frac{-s}{\mu^2}
\end{equation}
with $a_\mu=\alpha_s(\mu)/\pi$, we get the following expansion for the Adler function
\begin{equation}\label{eq:genexpAdler}
D_V^{(1+0)}(s)=\frac{N_c}{12\pi^2}\sum_{n=0}^\infty a_\mu^n 
  \sum_{k=1}^{n+1} k c_{n,k} \tilde\eta^{k-1}
\end{equation}
In the language of ref.~\cite{Le Diberder:1992te,Le Diberder:1992fr}, we obtain for the function arising in eq.~(\ref{eq:kngn})
\begin{equation}
\tilde{K}_n(\xi)=\sum_{k=1}^{n+1} k c_{n,k} (-\log\xi)^{k-1}
\end{equation}
But since the Adler function is independent of the renormalisation scale, as implemented in eq.~(\ref{eq:pertAdler}), we can take the derivative of eq.~(\ref{eq:genexpAdler}) with respect to $\log\mu$ and exploit the RGE to reexpress the derivative of $a_\mu$ in terms of $a_\mu$ itself. We obtain an expansion in powers of $a_\mu$ and $\eta$ which is identical to zero. This yields expressions for $c_{n,k\geq 2}$ from $c_{n,1}$. Since the values of $\beta_{n\ge 4}$ and $c_{n\geq 5,1}$ are unknown, the reconstruction of of $c_{n,k\geq 2}$ based on the RGE is only partial: the coefficients $c_{n,k}$ with $n\geq 6$ and $1 \leq k \leq n-4$ cannot be computed fully because their equations involve some of these unknown coefficients.

One can then perform the computation of the perturbative contribution to the tau width:
\begin{equation}\label{eq:an}
\delta^{(0)}=\sum_{n=1}^\infty \sum_{k=1}^n k c_{n,k} \frac{1}{2i \pi}
      \oint_{|x|=1} \frac{dx}{x} (1-x)^3 (1+x) \log^{k-1}\left(\frac{-M_\tau^2 x}{\mu^2}\right) a_\mu^n 
\end{equation}
in two different ways, either by performing FOPT or CIPT. In FOPT, we set $\mu=M_\tau$, leading to
\begin{equation}\label{eq:FOPTdelta0}
\delta^{(0)}_{FOPT}=\sum_{n=1}^\infty a_{M_\tau}^n \sum_{k=1}^n k c_{n,k} J_{k-1}
\end{equation}
with 
\begin{equation}
J_k=\frac{1}{2i \pi}
      \oint_{|x|=1} \frac{dx}{x} (1-x)^3 (1+x) \log^{k}(-x)
\end{equation}
In CIPT, we take $\mu^2=-M_\tau^2 x$ to get
\begin{equation}
\delta^{(0)}_{CIPT}=\sum_{n=1}^\infty c_{n,1} J^a_n
\end{equation}
with 
\begin{equation}
J^a_n=\frac{1}{2i \pi}
      \oint_{|x|=1} \frac{dx}{x} (1-x)^3 (1+x) a^n(-M_\tau^2 x)
\end{equation}

Since the two presentations are not obviously identical, it is interesting to determine if they are fully equivalent. In the case of CIPT, only the coefficients $c_{n,1}$ coefficients from eq.(\ref{eq:pertAdler}) are involved, so that the definitions are easily recognised as identical. In the case of FOPT, the situation is rather different, since we use the RGE at all orders to derive the coefficients $c_{n,k\geq 2}$ according to ref.~\cite{Beneke:2008ad}, whereas it is only exploited to up to a given order in the other references ($\alpha_s^5$ in ref.~\cite{Davier:2005xq}). The two results differ from each other through the higher-order terms that are kept or not. Indeed, the FOPT expression  contains terms up to a certain order in $\alpha_s(M_\tau)$ coming from:
\begin{itemize}
\item the perturbative expansion of $\hat{D}$, known up to an arbitrary order through the renormalon model
\item the RGE of $\alpha_s$, used to reexpress the integral in terms of a series in $\alpha_s(M_\tau)$, with coefficients known only up to $\beta_3$. 
\end{itemize}

The two previous definitions of FOPT can be rephrased in the following way:
\begin{itemize}
\item In ref.~\cite{Beneke:2008ad}, the RGE is used to expand $\alpha_s^n(s)$ for $n$ arbitrary, in powers of $\alpha_s^k(s_0)$ up to the required order in perturbation theory, setting all $\beta_{k\geq 4}=0$. Then these expansions are put in eq.~(\ref{eq:pertAdler}) to compute the FOPT value of $\delta^{(0)}$ at a given order in perturbation theory. This method is denoted FOPT(BJ) in the following.
\item In refs.~\cite{Davier:2005xq}, the expansion of $\alpha_s$, eq.~(\ref{eq:astaylor}) up to $O(\alpha_s^5)$ is plugged  into the series for $\delta^{(0)}$, eq.~(\ref{eq:FOPTdelta0}), the latter being cut at the required order in perturbation theory. We denote FOPT this way of dealing with higher orders.
\end{itemize}
The difference between the two methods comes from a different use of the RGE for orders above the known terms (i.e., once $\beta_{n\geq 4}$ is involved).
 For instance, for the $n=6$ contribution to $\delta^{(0)}$, we should use eq.~(\ref{eq:astaylor}) up to $\alpha_s^6$, involving $\beta_4$. This difference was already discussed in ref.~\cite{Davier:2008sk}, where only low orders were known and included. In this reference, we discussed a method called FOPT$^+$, which corresponds to
 the prescription denoted FOPT here, up to high-order coefficients in the perturbative expansion of $D$ (modeled here, but set to zero in FOPT$^+$). 
A second method, called FOPT$^{++}$ in that paper, is related to the prescription denoted FOPT(BJ) here (in ref.~\cite{Davier:2008sk}, we stopped the Taylor expansion at $\eta^{n\leq 5}$).

\begin{figure}[h]
\begin{center}
\includegraphics[width=12cm]{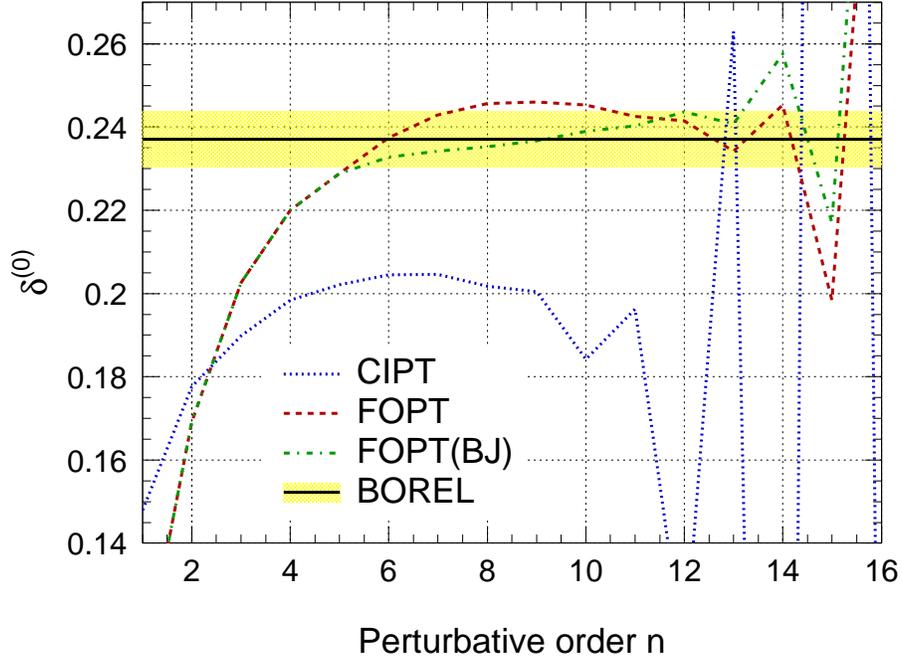}
\caption{Value of $\delta^{(0)}$ as a function of the order and type of perturbation theory, corresponding to Fig. 7 from ref.~\cite{Beneke:2008ad}.
FOPT and FOPT(BJ) indicate two different ways of dealing with Fixed-Order Perturbation Theory as one goes to higher orders in perturbation theory.}
\label{plotBJ}
\end{center}
\end{figure}

Following the methods presented in ref.~\cite{Beneke:2008ad}, we obtain the plot in Fig.~\ref{plotBJ} (in this paper, we will always use the illustrative value of the strong coupling constant $\alpha_s(M_\tau)=0.34$).
Let us emphasize that neither FOPT nor FOPT(BJ) is "the true" FOPT as soon as we look for contributions from orders higher than $n=5$, and they are both different from the complete value that would be derived by applying FOPT if the full RGE series were known: a piece is missing.

One can see that the exploitation of the RGE at higher orders has a significant impact on $\delta^{(0)}$. The difference between the two methods remains within our (conservative) error bars, but which exceeds the limited error bars chosen by ref.~\cite{Beneke:2008ad}. 
In the case of CIPT, one has also to decide up to which order of perturbation theory one should write eq.~(\ref{eq:astaylor}). However, the integration of RGE is performed by small-step integration, and thus it is much less dependent on the exact cut on the power of $\eta$ placed on eq.~(\ref{eq:astaylor})~\cite{Davier:2008sk}.

\section{Extension of the renormalon model}

Based on the renormalon model presented in sec.~\ref{sec:renormmodel},
ref.~\cite{Beneke:2008ad} fixed the renormalon residues $d_i$ in
eq.~(\ref{eq:ansatzBJ})
by requiring the model to reproduce the first orders of the perturbative series $c_{3,1}, c_{4,1}, c_{5,1}$ (the latter being set to $c_{5,1}=283$ following an ansatz concerning the geometrical growth of this coefficients). The argument stated in ref.~\cite{Beneke:2008ad} is that only a constant term $d_0^{PO}$ is actually needed to reproduce $c_{1,1}$ and $c_{2,1}$ (even though a linear term $d_1^{PO}$ is included, but turned out to be very small in the ansatz
of ref.~\cite{Beneke:2008ad}).

With this assumption, ref.~\cite{Beneke:2008ad}
shows that the balance between the two IR renormalons has a direct consequence on the discussion of CIPT/FOPT in their model:
\begin{itemize}
\item A dominance of the $d=6$ renormalon favours CIPT (in the sense that it yields a value close to the Borel sum of the series)
\item A dominance of the $d=4$ renormalon favours FOPT (in the sense that it yields a value close to the Borel sum of the series)
\end{itemize}

However, this procedure to fix the residues $d_i$ \emph{assumes} that the perturbative expansion eq.~(\ref{eq:perturbexp}) is valid exactly when one neglects the contributions coming from the coefficient $c_2$ and from $1/n^3$ remainders (since they are set to 0 in ref.~\cite{Beneke:2008ad}).
Actually, these contributions can be quite significant, compared to other contributions that are explictly included in the renormalon model in ref~\cite{Beneke:2008ad}. Indeed, Tables~\ref{tab:n3} and \ref{tab:c2} collect the relative contribution from $1/n^3$ and $c_2$ terms:
\begin{eqnarray}
E_p^{n,X}(a_Q)&=&\frac{\pi d_p^{X}}{p^{1+\gamma} \Gamma(1+\gamma)} 
  \Gamma(n+1+\gamma) \left(\pm \frac{\beta_0}{p}\right)^n a_Q^{n+1}\frac{1}{n^3}\\
F_p^{n,X}(a_Q)&=&\frac{\pi d_p^{X}}{p^{1+\gamma} \Gamma(1+\gamma)} 
   \Gamma(n+1+\gamma) \left(\pm \frac{\beta_0}{p}\right)^n a_Q^{n+1}
    \left(\frac{p}{\beta_0}\right)^2 \frac{1}{(n+\gamma)(n+\gamma-1)}
\end{eqnarray}
compared to the total contribution of a pole for a given order in perturbation theory as computed in ref~\cite{Beneke:2008ad}:
\begin{eqnarray}
D_p^{n,X}(a_Q)&=&\frac{\pi d_p^{X}}{p^{1+\gamma} \Gamma(1+\gamma)} 
  \Gamma(n+1+\gamma) \left(\pm \frac{\beta_0}{p}\right)^n a_Q^{n+1}\\
 && \qquad \times \left[1+\frac{p}{\beta_0} \frac{(b_1+c_1)}{n+\gamma}+\left(\frac{p}{\beta_0}\right)^2 \frac{(b_2+b_1c_1)}{(n+\gamma)(n+\gamma-1)}\right]
 \nonumber
\end{eqnarray}
In other words, the actual contribution to the perturbative expansion of the Adler series from a given pole at a given order would be
\begin{equation}
a_Q^{n+1}[D_p^{n,X} + z  E_p^{n,X} + c_2 F_p^{n,X}]
\end{equation}
$z$ and $c_2$ are unknown coefficients, in principle of order 1. One could neglect their presence at a given order $a_s^{n+1}$ of perturbation theory if $E_p^{n,X}/D_p^{n,X}$ and $F_p^{n,X}/D_p^{n,X}$ are small numbers (this is in particular assumed in ref~\cite{Beneke:2008ad} for $n=2,3,4$).

\begin{table}
\begin{center}
\begin{tabular}{ccccccccccccc}
\hline\noalign{\smallskip}
Pole & 1 & 2 & 3 & 4 & 5 & 6 & 7 & 8 & 9 & 10 & 11 & 12  \\
\hline\noalign{\smallskip}
UV &  36.1 &  11.7 &  3.7 &  
  1.6 &  0.8 &  0.5 &  0.3 &  
  0.2 &  0.1 &  0.1 &  0.1 & 
   0.1 \\   
IR2 & -377.4 &  46.2 &  7.5 &  
  2.5 &  1.2 &  0.6 &  0.4 &  
  0.2 &  0.2 &  0.1 &  0.1 &  
  0.1\\
IR3 & -106.3 &  -59.3 &  26.3 &  
  4.6 &  1.7 &  0.8 &  0.5 &  
  0.3 &  0.2 &  0.1 &  0.1 & 
   0.1\\
\hline\noalign{\smallskip}
 \end{tabular}
\caption{Relative contribution (in percent) $E_p^{n,X}/D_p^{n,X}$ from $1/n^3$ to a given order of perturbation theory for each pole of the renormalon model. The correction cannot be estimated for $n=0$.} \label{tab:n3}
\end{center}
\end{table}
   
\begin{table}
\begin{center}
\begin{tabular}{cccccccccccccc}  
\hline\noalign{\smallskip}
Pole & 0 & 1 & 2 & 3 & 4 & 5 & 6 & 7 & 8 & 9 & 10 & 11 & 12 \\
\hline\noalign{\smallskip}
UV &   36.4 &  28.1 &  6.7 &  2.8 &  
  1.5 &  0.9 &  0.6 &  0.4 &  
  0.3 &  0.3 &  0.2 &  0.2 & 0.1\\
IR2 & -29.7 &  -73.1 &  31.6 &  
  9.8 &  5.0 &  3.1 &  2.1 &  
  1.6 &  1.2 &  0.9 &  0.8 &  
  0.6 &  0.5\\ 
IR3 & -17.9 &  -23.7 &  -57.3 &  
  53.8 &  15.2 &  8.0 &  5.1 &  
  3.6 &  2.7 &  2.1 &  1.7 &  
  1.4 &  1.2\\
\hline\noalign{\smallskip}
 \end{tabular}
\caption{Relative contribution (in percent) $F_p^{n,X}/D_p^{n,X}$ from $c_2$ to a given order of perturbation theory for each pole of the renormalon mode.} \label{tab:c2}
\end{center}
\end{table}

One notices that the perturbative coefficient for $n\leq 3$ is significantly affected by both kinds of contribution ($E$ and $F$), and that the $IR_3$ pole is more affected by such corrections than $IR_2$. 
In addition, the value of $c_2$ can affect significantly the situation for $IR_3$ even for higher orders. Therefore, the model discussed in ref.~\cite{Beneke:2008ad} is likely to have its perturbative expansion affected by significant corrections at low orders -- where the term "low orders" includes the $O(a_s^3)$ term (i.e., $n=2$), which is used to determine the residues of the poles.

One can take into account the possibility of such large corrections at the lower orders
 by adding a quadratic term in $u$ to the polynomial part, which will contribute to the perturbative expansion of the Adler function at $O(a_s^3)$ :
\begin{equation}
B[\hat{D}](u)=B[\hat{D}_1^{UV}](u)+B[\hat{D}_2^{IR}](u)+B[\hat{D}_3^{IR}](u)
      +d_0^{PO}+d_1^{PO} u+d_2^{PO} u^2
\end{equation}

\begin{table*}[t] 
  \caption[.]{\label{tab:coeffs}
              Coefficients of the ansatz for different values of $d_2^{PO}$.}
\begin{center}
\setlength{\tabcolsep}{0.0pc}
\begin{tabular*}{\textwidth}{@{\extracolsep{\fill}}rrrrrrrrr} 
\hline\noalign{\smallskip}
         $d_0^{PO}$  & $d_1^{PO}$   &  $d_2^{PO}$  & $d_1^{UV}$ & $d_2^{IR}$   & $d_3^{IR}$  \\
\noalign{\smallskip}\hline\noalign{\smallskip}
	-9.116  &-3.834   &-1    & 0.028   & 7.82   &-176.8   \\
	-4.167  &-1.913  &-0.5  & 0.006 & 5.49  & -95.16  \\
	 0.781  & 0.008 & 0   & -0.0160 & 3.16   & -13.53 \\
	 3.255  & 0.968 & 0.25 & -0.026 & 1.99  & 27.29   \\
	 5.729  & 1.929  & 0.5  & -0.037 & 0.83 & 68.10   \\
	 10.68  & 3.850  & 1    &-0.059  &-1.50  & 149.7   \\
\noalign{\smallskip}\hline
\end{tabular*}
  \end{center}
\end{table*}

The choice of $d_2^{PO}$ can be seen converted into a guess on the value of $c_{6,1}$, with the corresponding equivalence:
\begin{center}
\begin{tabular}{ccccccc}
$d_2^{PO}$ &-1 & -0.5 & 0 & 0.25 & 0.5 & 1\\
\hline
$c_6$      & 1291  & 2283 & 3275 & 3771 & 4267  & 5259\\
\end{tabular}
\end{center}

\begin{figure}[h]
\begin{center}
\includegraphics[width=8cm]{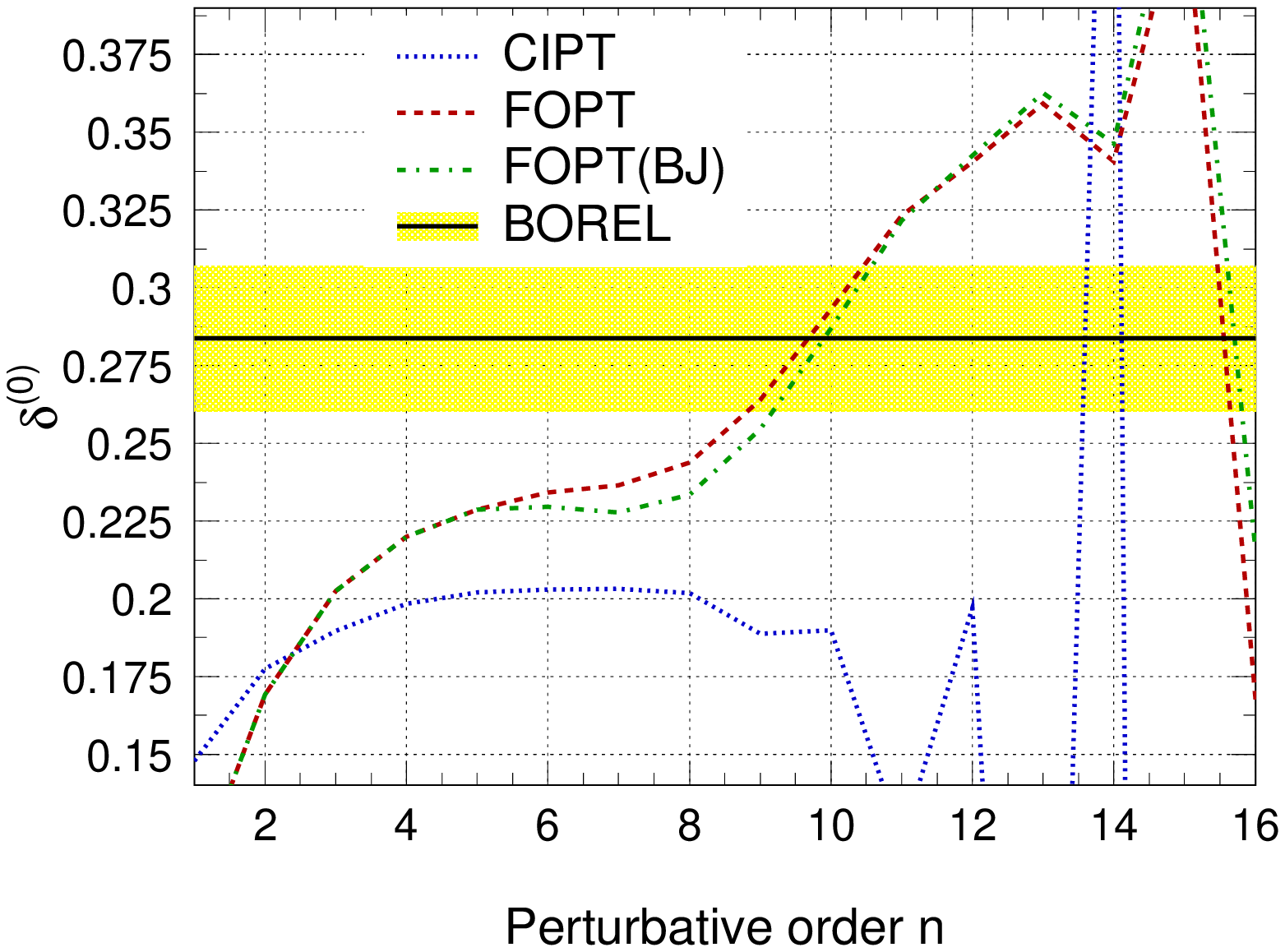}\includegraphics[width=8cm]{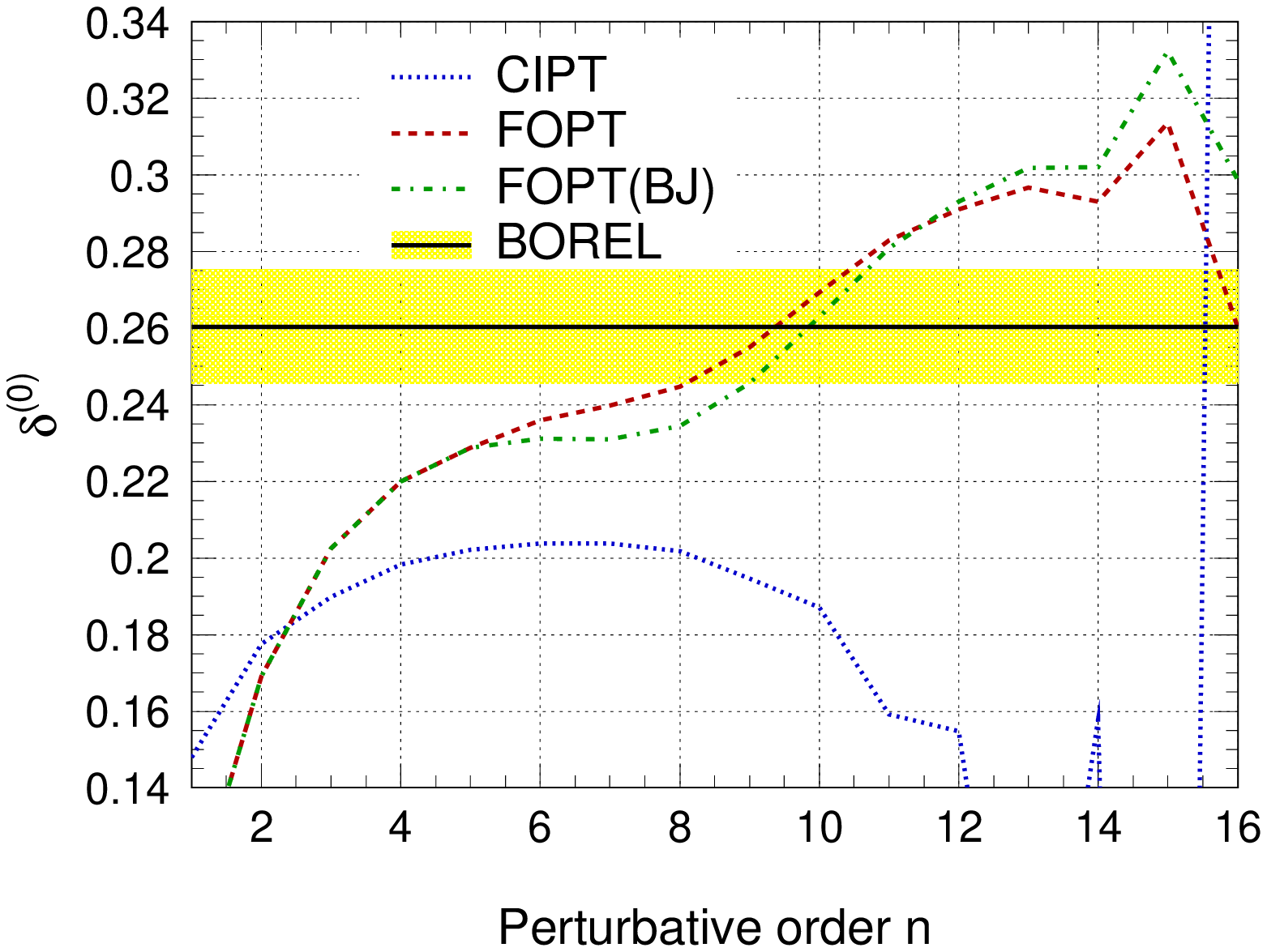}

\includegraphics[width=8cm]{standardb4.eps}\includegraphics[width=8cm]{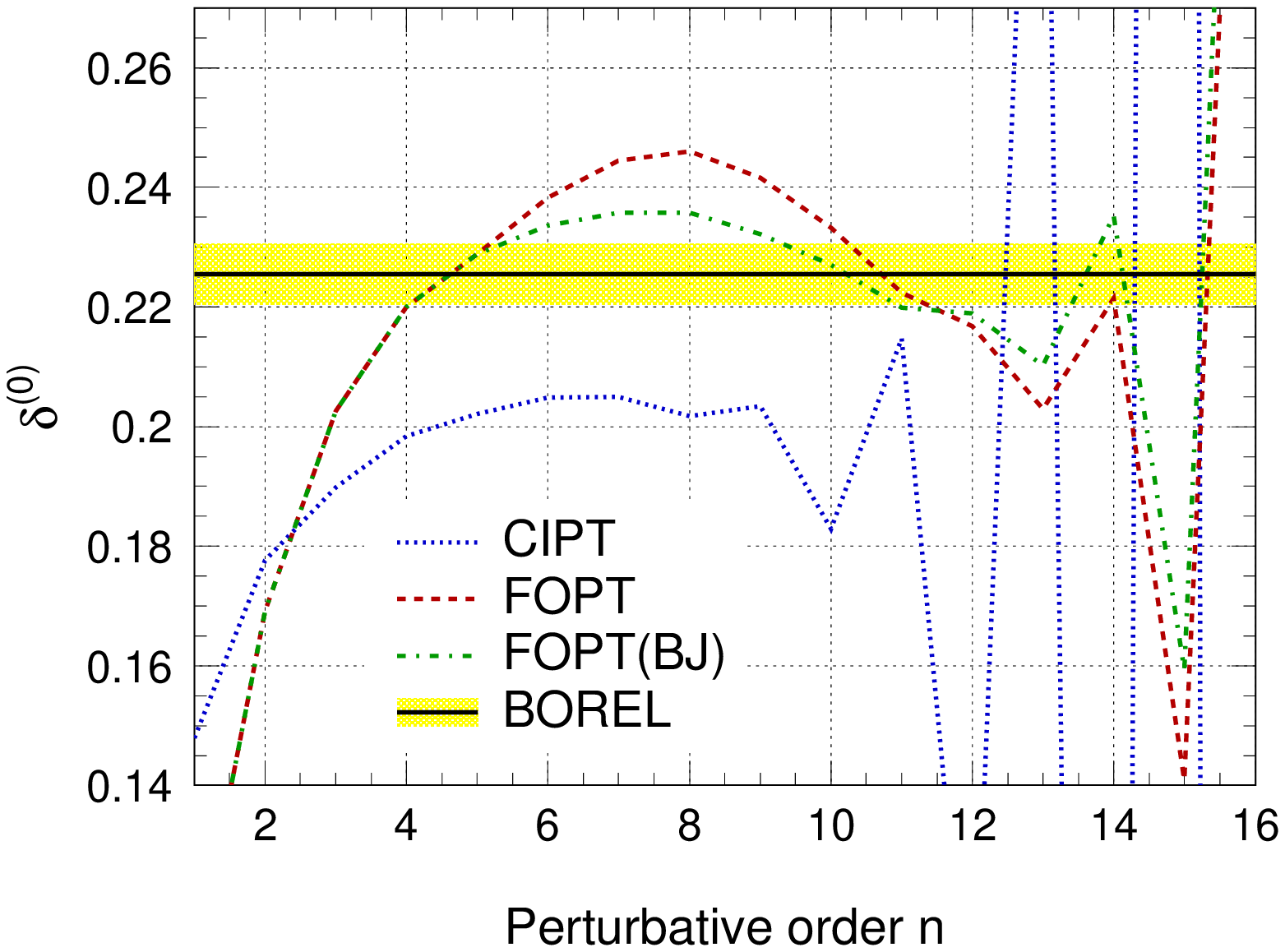}

\includegraphics[width=8cm]{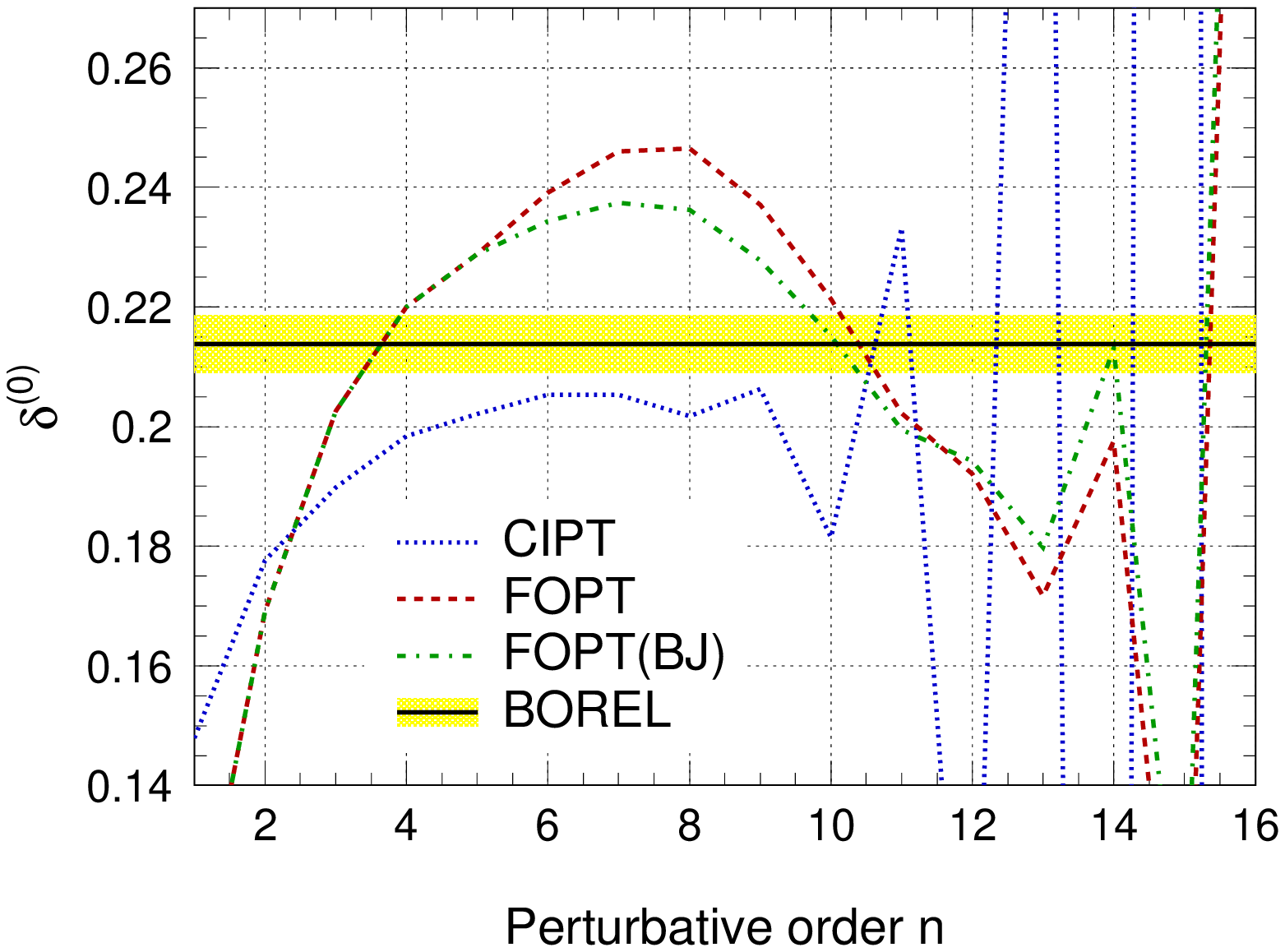}\includegraphics[width=8cm]{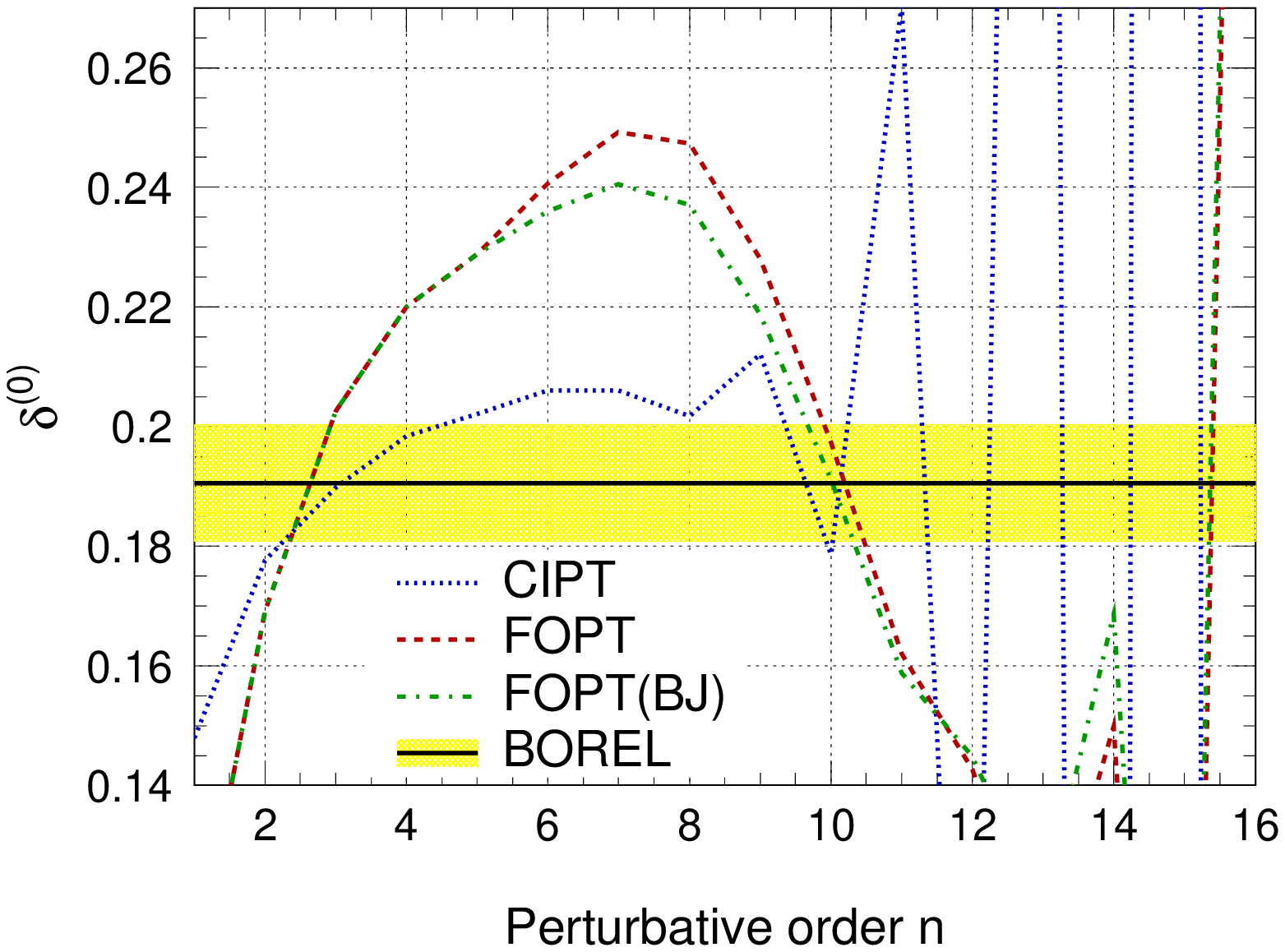}

\caption{
Value of $\delta^{(0)}$ as a function of the order and type of perturbation theory, corresponding to Fig. 7 from ref.~\cite{Beneke:2008ad}.
FOPT and FOPT(BJ) indicate two different ways of dealing with Fixed-Order Perturbation Theory as one goes to higher orders in perturbation theory.
The figures correspond to different values of the quadratic coefficient
$d_2^{PO}=-1,-0.5,0,0.25,0.5,1$ (from left to right and top to bottom)}
\label{2ndorderpol}
\end{center}
\end{figure}

The dependence of all results on $d_2^{PO}$ is linear, and one can perform exactly the same analysis as before, with the corresponding plots  in Fig.~\ref{2ndorderpol}.  As expected, $O(1)$ values of $d_2^{PO}$ are enough to change the balance between the IR renormalon poles, as shown in Table~\ref{tab:coeffs}.

One recovers the model in ref.~\cite{Beneke:2008ad} in the case where $d_2^{PO}=0$. The discussion of CIPT vs FOPT can be converted into a discussion on the value of $d_2^{PO}$.
The cases where $d_2^{IR}$ vanishes (for $d_2^{PO}$ approximately between 0.5 and 1)  correspond to cases where CIPT is preferred to FOPT if one wants an agreement with the value of $\delta^{(0)}$ from the Borel resummation. The cases where $d_3^{IR}$ vanishes (approximately $d_2^{PO}$ between 0 and 0.5) correspond to cases where FOPT is preferred to CIPT. 

\clearpage

Larger (positive or negative) values of $d_2^{PO}$ correspond to values where neither FOPT nor CIPT yield a plateau with a value in agreement with the Borel resummation.
CIPT yields a result that is stable, but in disagreement with the value obtained by Borel resummation, whereas FOPT yields an unstable result which sometimes crosses the Borel result. We see that the value of the Borel resummation depends significantly on the value of $d_2^{PO}$, as well as the uncertainty attached to it.


\begin{table*}
\caption[.]{Relative contribution (in \%) to the coefficients of the perturbative expansion of the Adler function for different values of $d_2^{PO}=-0.5,0,0.25,0.5$\label{tab:relcontd2PO}}
\begin{center}
\begin{tabular}{cccccccccccc}
\hline
$d_2^{PO}$ & Pole & $c_{4,1}$ & $c_{5,1}$ & $c_{6,1}$ & $c_{7,1}$ & $c_{8,1}$ & $c_{9,1}$ & 
    $c_{10,1}$ & $c_{11,1}$ & $c_{12,1}$ & $c_{13,1}$\\
\hline
 &UV$_1$ &  -3.8 & 6.1 & -8.9 & 9.2 & -15.9 &
  16.7 & -45.5 & 33.7 & -576.4 & 59.4\\
-0.5 &IR$_2$ & 174.4 & 236.1 &
  243.0 & 164.8 & 177.7 & 112.6 & 
  180.1 & 77.1 & 752.1 & 43.7\\
&IR$_3$ & -70.6 & -142.2 & -134.1 &
-74.1 & -61.7 & -29.4 & -34.6 &
-10.8 & -75.7 & -3.1\\
\hline
&UV$_1$ &  9.7 & -15.6 & 15.8 & -38.6 & 30.3 & -236.7 &  54.0 & 200.0 & 77.6 & 119.5\\
0 &IR$_2$ & 100.4 & 135.9 & 97.5 & 155.9 & 76.3 & 360 & 48.3 & -103.6 & 22.9 & -19.9\\
&IR$_3$ & -10.0 & -20.2 & -13.3 & -17.3 & -6.5 & -23.2 & -2.3 & 3.6 & -0.6 & 0.3\\
\hline
& UV$_1$ &  16.4 & -26.5 & 23.3 & -96.5 & 45.6 & 314.6 & 
  72.3 & 127.2 &  89.4 & 107.0 \\
0.25 & IR$_2$ & 63.3 & 85.7 & 53.4 & 145.0 & 42.7 & -178 & 24.1 &
  -24.5 & 9.8 & -6.6 \\
& IR$_3$ & 20.2 & 40.8 & 23.3 & 51.4 & 11.7 & -36.7 & 
  3.6 & -2.7 & 0.8 & -0.4 \\
\hline
& UV$_1$ &  3.1 & -37.4 & 29.0 & -258.7 & 57.8 & 159.4 & 84.2 & 110.4 & 95.4 & 102.5\\
0.5 & IR$_2$ & 26.3 & 35.6 & 19.6 & 114.6 & 15.9 & -26.6 &   8.3 & -6.3 & 3.1 & -1.9\\
& IR$_3$ & 50.5 & 101.8 & 51.3 & 244.1 & 26.3 & -32.9 & 
  7.5 & -4.1 & 1.5 & -0.6\\
\hline
\end{tabular}
\end{center}
\end{table*}

\section{Definition of pole dominance}

From the previous section, one can see that a seemingly small change in $d_2^{PO}$ has an important impact on higher orders and on the value obtained from the Borel resummation. It seems difficult to determine a priori which model is more relevant, and different criteria can be imagined.
A rather usual approach consists in assuming that the first IR pole should "dominate" over the following ones for the model to be reasonable. Let us remark first that such a requirement is by no means mandatory in the framework of the renormalon approach, in which the relative contributions form different poles is free, at least in principle.
For the time being, let us assume that such a dominance is indeed the sign of a good model. The next question is: what is the definition of this dominance in practice ?


\begin{table*}[h]
  \caption[.]{\label{table:relcoeff-0.5}
     Order-by-order results for $\delta^{(0)}$ FOPT(BJ) (top) and CIPT (bottom), for $d_2^{PO}$ = -0.5}
\begin{center}
\setlength{\tabcolsep}{0.0pc}
\begin{tabular*}{\textwidth}{@{\extracolsep{\fill}}lrrrrrrrrr} 
\hline\noalign{\smallskip}
 Order   &   $1$   &   $2$   &   $3$   &   $4$   &   $5$   &   $6$   \\
\noalign{\smallskip}\hline\hline\noalign{\smallskip}
UV$_1$	& -0.00682 & -0.01114 & -0.01277 & -0.01281 & -0.01219 & -0.01161 \\
IR$_2$	& -0.90568 & -1.28083 & -1.37745 & -1.33435 & -1.25036 & -1.17583 \\
IR$_3$	&  2.43770 &  3.58286 &  3.99478 &  3.99000 &  3.82183 &  3.64055\\
Pol	& -1.41698 & -2.12173 & -2.40198 & -2.42283 & -2.33050 &  -2.22195\\
\noalign{\smallskip}\hline\noalign{\smallskip}
Sum FOPT(BJ)	& 0.10822 & 0.16916 & 0.20258 & 0.22001 & 0.22878 & 0.23116 \\
\noalign{\smallskip}\hline\hline\noalign{\smallskip}
UV$_1$	&  -0.00932 & -0.01194 & -0.01141 & -0.01174
  & -0.01150 & -0.01165\\
IR$_2$	&    -1.23768 & -1.27781 & -1.26035 & -1.24537 &
-1.23643 & -1.23238\\
IR$_3$	&    3.33131 & 3.64857 & 3.67319 & 3.66712 & 3.66174 & 3.65950\\
Pol	&    -1.93642 & -2.18124 & -2.21166 & -2.21166 &
-2.21166 & -2.21166\\
\noalign {\smallskip}\hline\noalign{\smallskip}
Sum (CIPT)	&  0.14790 & 0.17758 & 0.18977 & 0.19836 &
0.20214 & 0.20381\\
\noalign {\smallskip}\hline\hline\noalign{\smallskip}
Borel sum & 0.26036\\
\noalign{\smallskip}\hline

\end{tabular*}
  \end{center}
\end{table*}

\begin{table*}[h]
  \caption[.]{\label{table:relcoeff0}
         Order-by-order results for $\delta^{(0)}$ FOPT(BJ) (top) and CIPT (bottom), for $d_2^{PO}$ = 0.}
\begin{center}
\setlength{\tabcolsep}{0.0pc}
\begin{tabular*}{\textwidth}{@{\extracolsep{\fill}}lrrrrrrrrr} 
\hline\noalign{\smallskip}
Order   &   $1$   &   $2$   &   $3$   &   $4$   &   $5$   &   $6$   \\
\noalign{\smallskip}\hline\hline\noalign{\smallskip}
UV$_1$	& 0.01734 & 0.02836 & 0.03248 & 0.032583 & 0.031007 & 0.029536 \\
IR$_2$	&-0.52120 &-0.73709 &-0.79269 &-0.767886 &-0.719556 &-0.676666 \\
IR$_3$	& 0.34660 & 0.50942 & 0.56798 & 0.567304 & 0.543394 & 0.517619 \\
Pol	& 0.26548 & 0.36847 & 0.39480 & 0.388012 & 0.373940 & 0.362262 \\
\noalign{\smallskip}\hline\noalign{\smallskip}
Sum FOPT(BJ)	& 0.10822 & 0.16916 & 0.20258 & 0.22001 & 0.22878 & 0.23275 \\
\noalign{\smallskip}\hline\hline\noalign{\smallskip}
UV$_1$	&   0.02370 & 0.03039 & 0.02903 & 0.02986 &
 0.02927 &  0.02965\\
IR$_2$	&  -0.71226 &-0.73534 &-0.72530 &-0.71668 &
-0.71154 & -0.70920\\
IR$_3$	&   0.47365 & 0.51876 & 0.52226 & 0.52140 &
 0.52063 &  0.52031\\
Pol	&   0.36280 & 0.36378 & 0.36378 & 0.36378 &
0.36378 & 0.36378\\
\noalign{\smallskip}\hline\noalign{\smallskip}
Sum (CIPT)	&   0.14790 &  0.17758 & 0.18977 & 0.19836 &
0.20214 & 0.20453\\
\noalign {\smallskip}\hline\hline\noalign{\smallskip}
Borel sum & 0.23709\\
\noalign{\smallskip}\hline
\end{tabular*}
  \end{center}
\end{table*}

\begin{table*}[h]
  \caption[.]{\label{table:relcoeff0.25}
        Order-by-order results for $\delta^{(0)}$ FOPT(BJ) (top) and CIPT (bottom), for $d_2^{PO}$ = 0.25}
\begin{center}
\setlength{\tabcolsep}{0.0pc}
\begin{tabular*}{\textwidth}{@{\extracolsep{\fill}}lrrrrrrrrr} 
\hline\noalign{\smallskip}
Order   &   $1$   &   $2$   &   $3$   &   $4$   &   $5$   &   $6$   \\
\noalign{\smallskip}\hline\hline\noalign{\smallskip}
UV$_1$	&  0.02943 &  0.04811 &  0.05511 &  0.05528 &  0.05260 &  0.05011 \\
IR$_2$	& -0.32896 & -0.46522 & -0.50031 & -0.48466 & -0.45415 & -0.42708 \\
IR$_3$	& -0.69896 & -1.02730 & -1.14541 & -1.14404 & -1.09582 & -1.04385 \\
Pol	&  1.10671 &  1.61357 &  1.78311 &  1.77170 &  1.70077 &  1.63356 \\
\noalign{\smallskip}\hline\noalign{\smallskip}
Sum	&  0.10822 &  0.16916 &  0.20258 &  0.22001 &  0.22878 &  0.23355 \\
\noalign{\smallskip}\hline\hline\noalign{\smallskip}
UV$_1$	&  0.04021 &  0.05156 &  0.04925 &  0.05066 &  0.04966 &  0.05030\\
IR$_2$	& -0.44954 & -0.46412 & -0.45778 & -0.45234 & -0.44909 & -0.44762\\
IR$_3$	& -0.95518 & -1.04615 & -1.05320 & -1.05146 & -1.04992 & -1.04928\\
Pol	&  1.51241 &  1.63629 &  1.65150 &  1.65150 &  1.65150 &  1.65150\\
\noalign{\smallskip}\hline\hline\noalign{\smallskip}
Sum	&   0.147898 & 0.17758 & 0.189766 & 0.198358 & 0.202144 & 0.204897\\
\noalign {\smallskip}\hline\hline\noalign{\smallskip}
Borel sum & 0.22546\\
\noalign{\smallskip}\hline
\end{tabular*}
  \end{center}
\end{table*}

\begin{table*}[h]
  \caption[.]{\label{table:relcoeff0.5}
       Order-by-order results for $\delta^{(0)}$ FOPT(BJ) (top) and CIPT (bottom), for $d_2^{PO}$ = 0.5}
\begin{center}
\setlength{\tabcolsep}{0.0pc}
\begin{tabular*}{\textwidth}{@{\extracolsep{\fill}}lrrrrrrrrr} 
\hline\noalign{\smallskip}
Order   &   $1$   &   $2$   &   $3$   &   $4$   &   $5$   &   $6$   \\
\noalign{\smallskip}\hline\hline\noalign{\smallskip}
UV$_1$	&  0.04151 &  0.06787 &  0.07774 &  0.07797 &  0.07420 &  0.07068 \\
IR$_2$	& -0.13672 & -0.19335 & -0.20793 & -0.20143 & -0.18875 & -0.17750 \\
IR$_3$	& -1.74451 & -2.56403 & -2.85881 & -2.85539 & -2.73504 & -2.60531 \\
Pol	&  1.94795 &  2.85867 &  3.17143 &  3.15538 &  3.02761 &  2.90485 \\
\noalign{\smallskip}\hline\noalign{\smallskip}
Sum FOPT(BJ) & 0.10822 & 0.16916 & 0.20258 & 0.22001 & 0.22878 & 0.23435 \\
\noalign{\smallskip}\hline\hline\noalign{\smallskip}
UV$_1$	&    0.05672 &  0.07273 &  0.06947 &  0.07146 &  0.07004 & 0.07095\\
IR$_2$	&   -0.18683 & -0.19289 & -0.19026 & -0.18799 & -0.18664 &-0.18603\\
IR$_3$	&   -2.38401 & -2.61105 & -2.62867 & -2.62433 & 
-2.62047 &  -2.61887\\
Pol	&    2.66202 &  2.90880 &  2.93922 &  2.93922 & 2.93922 & 2.93922\\
\noalign{\smallskip}\hline\noalign{\smallskip}
Sum CIPT &   0.14790 & 0.17758 & 0.18977 & 0.19836 & 0.20214 & 0.20526\\
\noalign{\smallskip}\hline\hline\noalign{\smallskip}
Borel sum & 0.21382\\
\noalign{\smallskip}\hline
\end{tabular*}
  \end{center}
\end{table*}

A first definition, chosen in ref.~\cite{Beneke:2008ad}, consists in considering the contribution of the different poles to the expansion of the Adler function at intermediate orders (for $n$ between 4 and 8). A dominant pole gives the largest contribution to each coefficient of the perturbative series for intermediate orders. In ref.~\cite{Beneke:2008ad}, the case $d_2^{PO}=0$ is considered as interesting because the relative contribution to a given $c_n,1$ from IR$_2$ is larger than IR$_3$.  This is recalled in Table~\ref{tab:relcontd2PO}. Let us remark that already for $n=7$, large cancellations occur between the UV pole and the first IR pole. If one now considers different values of $d_2^{PO}$, one notices that $d_2^{P0}=0.25$ is also in good agreement with this criterion. In this case, there is no plateau for FOPT at intermediate orders in perturbation theory, and the outcome of the Borel resummation lies between FOPT and CIPT. Larger positive values of $d_2^{PO}$ would be disfavoured according to this definition of dominance.
On the other hand, negative values of $d_2^{PO}$, in particular large ones, fulfill this definition, as can be seen for instance from Table.~\ref{tab:relcontd2PO}.

But one may wonder whether the relative contribution to intermediate orders of $D$ the best way of determining whether one element of the model or another yields the most significant contribution to $\delta^{(0)}$. Indeed, the intermediate orders contribute only mildly to the actual value of $\delta^{(0)}$, since they are multiplied by higher and higher powers of $\alpha_s$.
It seems rather natural to break down the contribution to $\delta^{(0)}$ into the contributions from the UV pole, the two IR poles and the polynomial term.

An alternative definition of the dominance from one pole would correspond to providing most of the contribution to $\delta^{(0)}$. Looking at Table~\ref{table:relcoeff0}, we see that in the case of $\delta^{(0)}$ for $d_2^{PO}=0$, and contrary to our intuition, the so-called dominant IR$_2$ pole yields a contribution that is almost canceled by the so-called subdominant IR$_3$ pole -- the contributions from the two IR poles and the polynomial part being of the same order of magnitude.
When one increases $d_2^{PO}$ (see Tables~\ref{table:relcoeff0.25} and \ref{table:relcoeff0.5}), one can notice that the contribution from the IR$_3$ pole grows and is canceled by the polynomial part. The value $d_2^{PO}\simeq 0.08$ corresponds to the extreme situation where the residue of the $IR_3$ pole vanishes (and thus this pole does not contribute) and the $IR_2$ pole saturates the contribution from IR renormalons. It seems fair to require all the various contributions (individual "poles" and polynomial term) to yield contributions of the same size, which is the case for $-0.5 \leq d_2^{PO} \leq 0.25$ (see Tables~\ref{table:relcoeff-0.5}-\ref{table:relcoeff0.25}), corresponding to a rather wide range of behaviours of FOPT/CIPT/Borel sum.

\section{Anomalous dimension for the operator of dimension 6}

As can be seen from eq.~(\ref{eq:cutpole}), the term of renormalon "pole" is slightly misleading, since one expects radiative corrections to turn these poles into cuts in the Borel plane. One can relate both types of renormalons to QCD operators. In particular, the presence of IR renormalons mirrors the existence of condensates in the OPE of the correlator under scrutiny~\cite{Beneke:1998ui}. 
In particular, the anomalous dimensions of the latter are used to fix some unknown parameters of the model. Ref.~\cite{Beneke:2008ad} considers:
\begin{itemize}
\item for $UV_1$, a vanishing anomalous dimension for a single "effective" condensate
\item for $IR_2$, the anomalous dimension corresponding to the gluon condensate [eq.~(5.13) in this reference]
\item for $IR_3$, a vanishing anomalous dimension for a single "effective" condensate
\end{itemize}

As explained in ref.~\cite{Beneke:1998ui}, the structure of the cuts becomes rather involved once the full set of operators are considered. $IR_2$ is rather simple since only one operator is involved by dimensional arguments: this renormalon is linked to $d=4$ operators, and only the gluon condensate is involved (neither the identity operator nor the quark condensate since we work in the massless limit). On the other hand, both $UV_1$ and $IR_3$ are related to dimension-6 operators, namely:
\begin{eqnarray}
&& (\bar\psi\gamma_\mu\psi)(\bar\psi\gamma^\mu\psi)\qquad (\bar\psi\gamma_\mu\gamma_5\psi)(\bar\psi\gamma^\mu\gamma_5\psi)\\
&& (\bar\psi\gamma_\mu T^A\psi)(\bar\psi\gamma^\mu T^A\psi)\qquad (\bar\psi\gamma_\mu\gamma_5T^A\psi)(\bar\psi\gamma^\mu\gamma_5T^A\psi)\qquad
f_{ABC} G_{\mu\nu}^A G^{\nu\ B}_\rho G^{\rho\mu C}
\end{eqnarray}

In ref.~\cite{Beneke:1998ui}, the anomalous dimensions of $d=6$ operators were reconsidered. After diagonalising the RG mixing matrix, the diagonal operators were shown to have  anomalous dimensions at one loop of the form 
\begin{equation}
\gamma^{(1)}_{O_6^i}=\beta_1 \lambda_i \qquad \lambda_i=\{0.379,0.126,-0.332,-0.753,0\}
\end{equation}
These five contributions should a priori be included individually in the renormalon model, rather than through a single "effective" condensate of unclear anomalous dimension, and actually set to zero. Since the $d=4$ operator is described with its correct anomalous dimension and since we compare $d=4$ and $d=6$ renormalons, it seems fair to treat both
renormalons on the same footing.

As can be seen from the perturbative expansion eq.~(\ref{eq:perturbexp}), the large-order behaviour of a "pole" is:
\begin{equation}
D_V^{(1+0)}(Q^2)=\sum_{n=0}^\infty r_n a_Q^n 
 \qquad r_n \sim_{n \to\infty} \left(\frac{a_Q \beta_0}{p}\right)^n n!\ n^\gamma
\end{equation}
We see that the larger the anomalous dimension $\gamma$, the more dominant the pole at large $n$. But conversely, at smaller $n$, the operators with smaller anomalous dimensions compete (and can even be more significant) than the pole with the largest $n$. Therefore, once the proper cut structure of the second IR pole is taken into account, it becomes necessary but difficult to fix the relative "strengths" of the cuts from the different operators from the lowest orders of perturbation theory.

\begin{figure}
\begin{center}
\includegraphics[width=8cm]{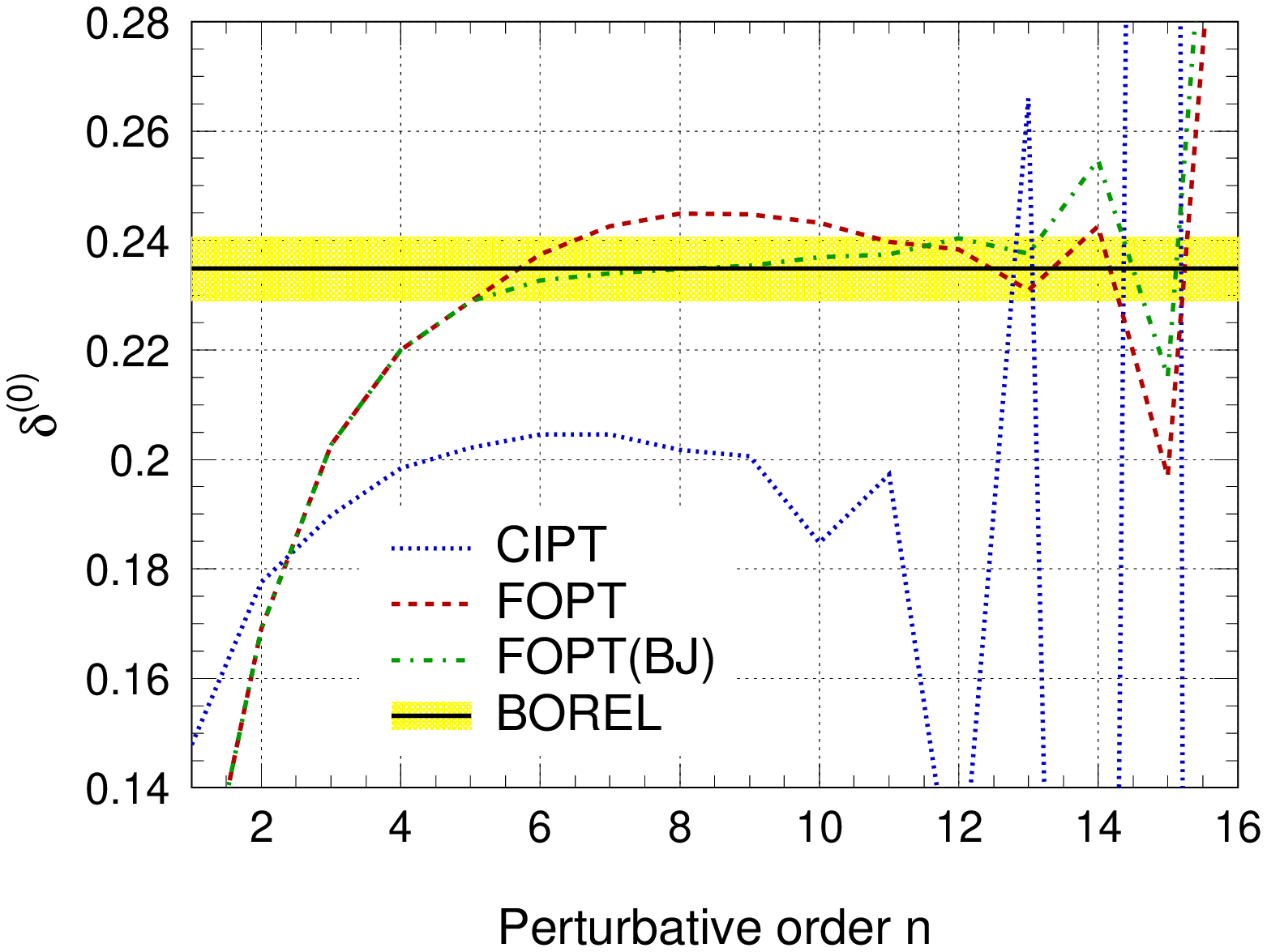}\includegraphics[width=8cm]{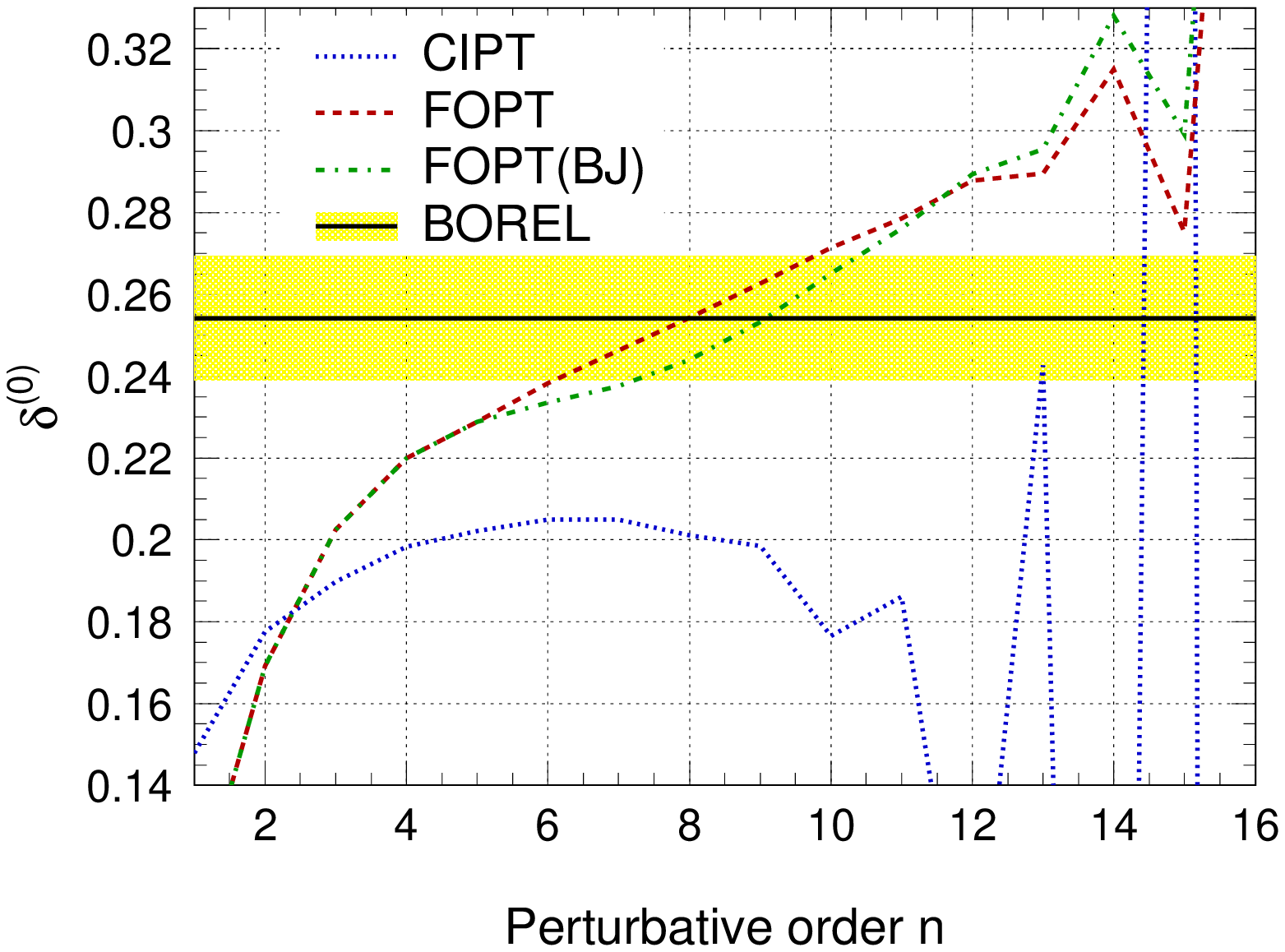}
\caption{Figures for different values of the anomalous dimension of the effective $d=6$ condensate:  $\gamma^{(1)}_{O_6}=\beta_1 \lambda$ with $\lambda=0.379$ (left) and $-0.753$ (right).}
\label{anomdim6}
\end{center}
\end{figure}

As an illustration of the role of anomalous dimensions in the discussion, we set $\gamma^{(1)}_{O_6}=\beta_1 \lambda$ with $\lambda=-0.753$ and $0.379$ (rather than 0 in the original model), and follow the same procedure as in ref.~\cite{Beneke:2008ad} to obtain the two plots in fig.~\ref{anomdim6}, indicating a rather wide range of behaviours for FOPT, depending on the choice of anomalous dimensions (CIPT on the other hand remains very stable). In addition, the resummed values for $\delta^{(0)}$ can vary from 0.23455 to 0.25553 if one changes this single parameter.

We can extend the model of ref.~\cite{Beneke:2008ad} by assuming a value for $c_{6,1}$ and splitting the second IR pole into two different poles, with different anomalous dimensions:
\begin{equation}
B[\hat{D}](u)=B[\hat{D}_1^{UV}](u)+B[\hat{D}_2^{IR}](u)+B[\hat{D}_{3a}^{IR}](u)+B[\hat{D}_{3b}^{IR}](u)
      +d_0^{PO}+d_1^{PO} u
\end{equation}
We denote $IR_{3a}$ for the cut with $\lambda_a=0.379$ and $IR_{3b}$ with
$\lambda_b=-0.753$. If we vary $c_{6,1}$ between 2283 and 4267 (values corresponding $d_2^{PO}=\pm 0.5$ in the previous model) and look at the relative contribution of each pole to a given order of $D$ (Table~\ref{tab:relcontc6}), we see that for values above 3000, one has a dominance of $IR_2$, with a significant cancellation of $IR_{3a}$ and a growing contribution from $IR_{3b}$. On the other hand, below 3000, $IR_{3a}$ takes over $IR_2$, whereas $IR_{3b}$ remains small. 
If we now consider the contributions to $\delta^{(0)}$ (Tables~\ref{tab:delta0c63275}-\ref{tab:delta0c64267}) and if we require them to be of similar order, values of  $c_{6,1}$ above 3000 are acceptable. In this case, it is quite interesting to notice that the breakdown in terms of the contributions from different poles is very different, as well as the relative contributions
to given orders of perturbation theory. Even though the value of $\delta^{(0)}$ is quite similar \emph{in fine}, the agreement with FOPT and CIPT depends quite strongly on the value chosen for $c_{6,1}$.

\begin{table*}
\caption[.]{Relative contribution (in \%) to the coefficients of the perturbative expansion of the Adler function for different values of $c_{6,1}$, for a model with two IR poles for $u=3$.
\label{tab:relcontc6}}
\begin{center}
\begin{tabular}{cccccccccccc}
\hline
$c_{6,1}$ & Pole & $c_{4,1}$ & $c_{5,1}$ & $c_{6,1}$ & $c_{7,1}$ & $c_{8,1}$ & $c_{9,1}$ & 
    $c_{10,1}$ & $c_{11,1}$ & $c_{12,1}$ & $c_{13,1}$\\
\hline
3275 & UV$_1$ &  9.7 & -15.7 & 15.9 & -38.9 &
  30.4 & -241.3 & 54.1 & 199.1 & 77.7 & 119.4\\
& IR$_2$ & 100. & 136.6 &
  98.0 & 157.3 & 76.7 & 367.5 & 
  48.5 & -103.3 &  23.0 & -19.8\\
& IR$_{3a}$ & -9.4 & -12.8 & -8.2 & -11.1 & -4.4 & -16.7 &
-1.7 & 2.8 & -0.5 & 0.3\\
& IR$_{3b}$ & -1.2 & -8.0 & -5.6 & -7.3 &
  -2.7 & -9.4 & -0.9 & 
  1.3 &-0.2 & 0.1\\
\hline
2283 & UV$_1$ &   5.8 & -9.4 & 13.6 &  -59.6 &
56.9 & 95.6 & 139.3 & 77.8 & 
128.0 & 87.1 \\
& IR$_2$ &    -81.4& -110.3&  -113.5 & -324.6 &
-193.3 & -196.0 & -168.2 &  54.3& 
-50.9 & 19.5 \\
& IR$_{3a}$ &    179.4 & 244.9 & 225.2 & 542.9 & 262.9 &
-211.3 & 140.9 & -34.9 & 24.7 & 
-7.1 \\
& IR$_{3b}$ &   -3.8 & -25.3 & -25.4 &  -58.7 & 
-26.4 & 19.6 & -12.1 & 2.8 & 
-1.8 & 0.5 \\
\hline
4267 &UV$_1$ &  13.6 & -22.0 & 17.0 & -33.8 &
25.3 & -96.2 & 42.9 & 599.3 & 66.5 & 142.0 \\
&IR$_2$ &  283.1 & 383.2 & 211.0&  274.4 & 128.2& 293.7 & 77.1 & -623.0 & 39.4 &
-47.4 \\
&IR$_{3a}$ &  -198.0 & -270.4 & -133.0 & -145.8 &
-55.4 & -100.5 & -20.5 & 127.0 &
-6.1 & 5.5 \\
&IR$_{3b}$ &  1.4 & 9.1 & 4.9 & 5.2 & 1.8 &
3.1 & 0.6 & -3.3 & 0.1 &
-0.1 \\
\hline
\end{tabular}
\end{center}
\end{table*}

\begin{table*}[h]
  \caption[.]{\label{tab:delta0c63275}
              Results for the FOPT(BJ) integral, for $c_{6,1}=3275$ in a model with two IR poles for $u=3$.}
\begin{center}
\setlength{\tabcolsep}{0.0pc}
\begin{tabular*}{\textwidth}{@{\extracolsep{\fill}}lrrrrrrrrr} 
\hline\noalign{\smallskip}
Order  &   $1$   &   $2$   &   $3$   &   $4$   &   $5$   &   $6$   \\
\noalign{\smallskip}\hline\noalign{\smallskip}
UV$_1$	  &   0.01740 &  0.02846 &  0.03259 & 0.03269 &  0.03111 &  0.02964 \\
IR$_2$	  &  -0.52346 & -0.74029 & -0.79613 &-0.77122 & -0.72268 & -0.67960\\
IR$_{3a}$ &   0.03761 &  0.05553 &  0.06153 & 0.06031 &  0.05618 &  0.05208\\
IR$_{3b}$ &   0.41127 &  0.59763 &  0.66170 & 0.65995 &  0.63438 &  0.60771\\
Pol	&     0.16540 &  0.22783 &  0.24289 & 0.23829 &  0.22979 &  0.22293\\
\noalign{\smallskip}\hline\noalign{\smallskip}
Sum FOPT(BJ)	&     0.10822 &  0.16916 &  0.20258 & 0.22001 &  0.22878 & 0.23275\\
\noalign{\smallskip}\hline\hline\noalign{\smallskip}
Borel sum &   0.23655 \\
\noalign{\smallskip}\hline
\end{tabular*}
  \end{center}
\end{table*}

\begin{table*}[h]
  \caption[.]{
              Results for the FOPT(BJ) integral, for $c_{6,1}=2283$ in a model with two IR poles for $u=3$.}
\begin{center}
\setlength{\tabcolsep}{0.0pc}
\begin{tabular*}{\textwidth}{@{\extracolsep{\fill}}lrrrrrrrrr} 
\hline\noalign{\smallskip}
Order   &   $1$   &   $2$   &   $3$   &   $4$   &   $5$   &   $6$   \\
\noalign{\smallskip}\hline\noalign{\smallskip}
UV$_1$	&   0.01044 & 0.01707 & 0.01955 & 0.01961 & 0.01866 & 0.01778 \\
IR$_2$	&   0.42298 & 0.59818 & 0.64330 & 0.62317 & 0.58395 & 0.54915 \\
IR$_{3a}$ &-0.72479 &-1.07018 &-1.18572 &-1.16230 &-1.08269 &-1.00363\\
IR$_{3b}$ & 1.29039 & 1.87510 & 2.07611 & 2.07060 & 1.99040 & 1.90671\\
Pol	&  -0.89078 &-1.25101 &-1.35066 &-1.33107 &-1.28154 &-1.23885\\
\noalign{\smallskip}\hline\noalign{\smallskip}
Sum FOPT(BJ)	&   0.10822 & 0.16916 & 0.20258& 0.220014 & 0.22878 & 0.23116\\
\noalign{\smallskip}\hline\hline\noalign{\smallskip}
Borel sum & 0.19598\\
\noalign{\smallskip}\hline
\end{tabular*}
  \end{center}
\end{table*}

\begin{table*}[h]
  \caption[.]{\label{tab:delta0c64267}
              Results for the FOPT(BJ) integral, for $c_{6,1}=4267$ in a model with two IR poles for $u=3$.}
\begin{center}
\setlength{\tabcolsep}{0.0pc}
\begin{tabular*}{\textwidth}{@{\extracolsep{\fill}}lrrrrrrrrr} 
\hline\noalign{\smallskip}
 Order   &   $1$   &   $2$   &   $3$   &   $4$   &   $5$   &   $6$   \\
\noalign{\smallskip}\hline\noalign{\smallskip}
UV$_1$	&   0.02437 &  0.03984 & 0.04563&  0.04577 & 0.04356 & 0.04149\\
IR$_2$	&  -1.46990 & -2.07876 &-2.23557& -2.16562 &-2.02932 &-1.90836\\
IR$_{3a}$ &  0.80001 &  1.18125 & 1.30878&  1.28293 & 1.19506 & 1.10779\\
IR$_{3b}$ & -0.46784 & -0.67983 &-0.75271& -0.75071 &-0.72163 &-0.69129\\
Pol	&   1.22158 &  1.70666 & 1.83645&  1.80764 & 1.74112 & 1.68471\\
\noalign{\smallskip}\hline\noalign{\smallskip}
Sum FOPT(BJ)	&   0.10822 &  0.16916 & 0.20258&  0.22001& 0.22878 &  0.23434\\
\noalign{\smallskip}\hline\hline\noalign{\smallskip}
Borel sum & 0.27712\\
\noalign{\smallskip}\hline
\end{tabular*}
  \end{center}
\end{table*}

\section{Other moments}

\begin{figure}[h]
\begin{center}
\includegraphics[width=8cm]{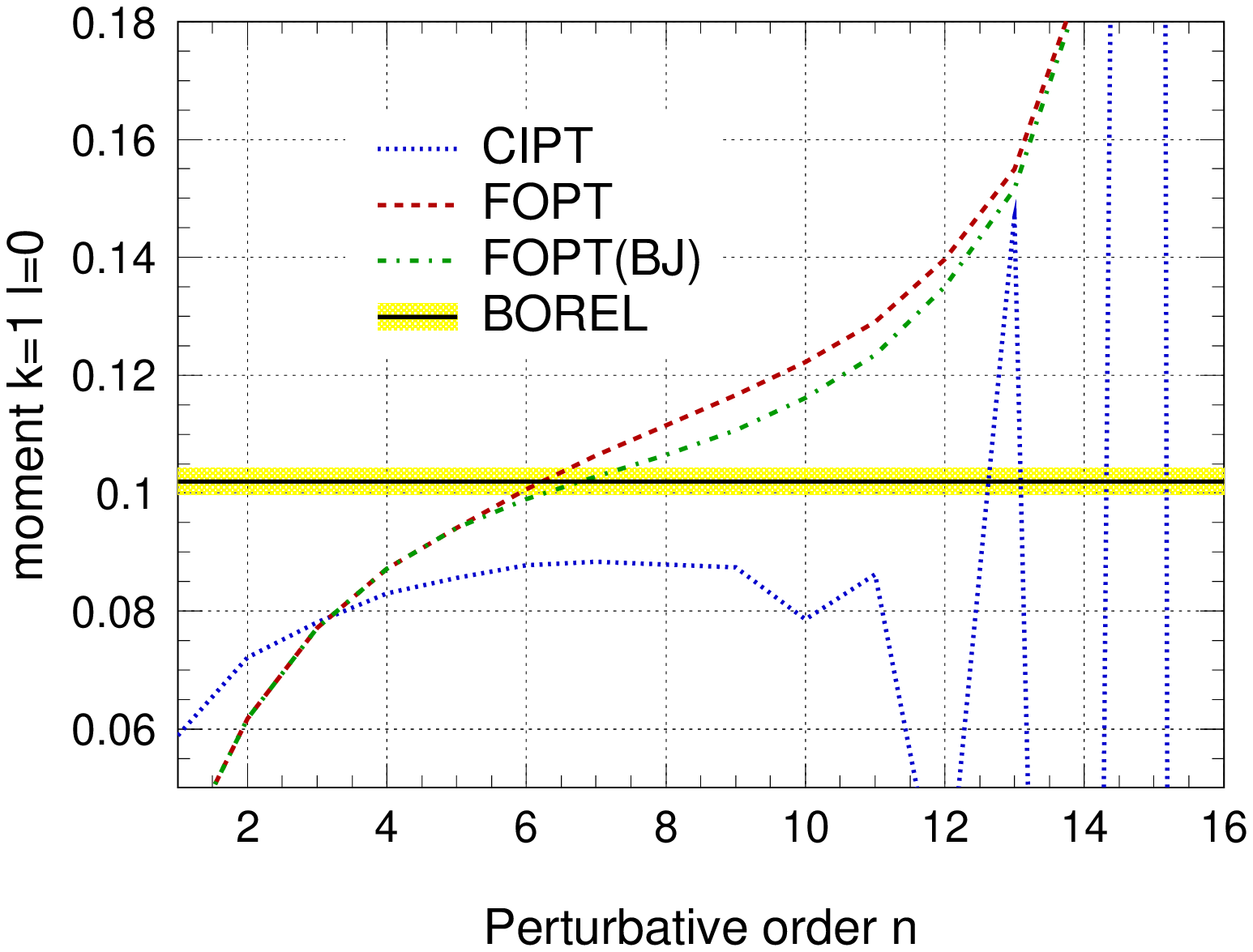}\includegraphics[width=8cm]{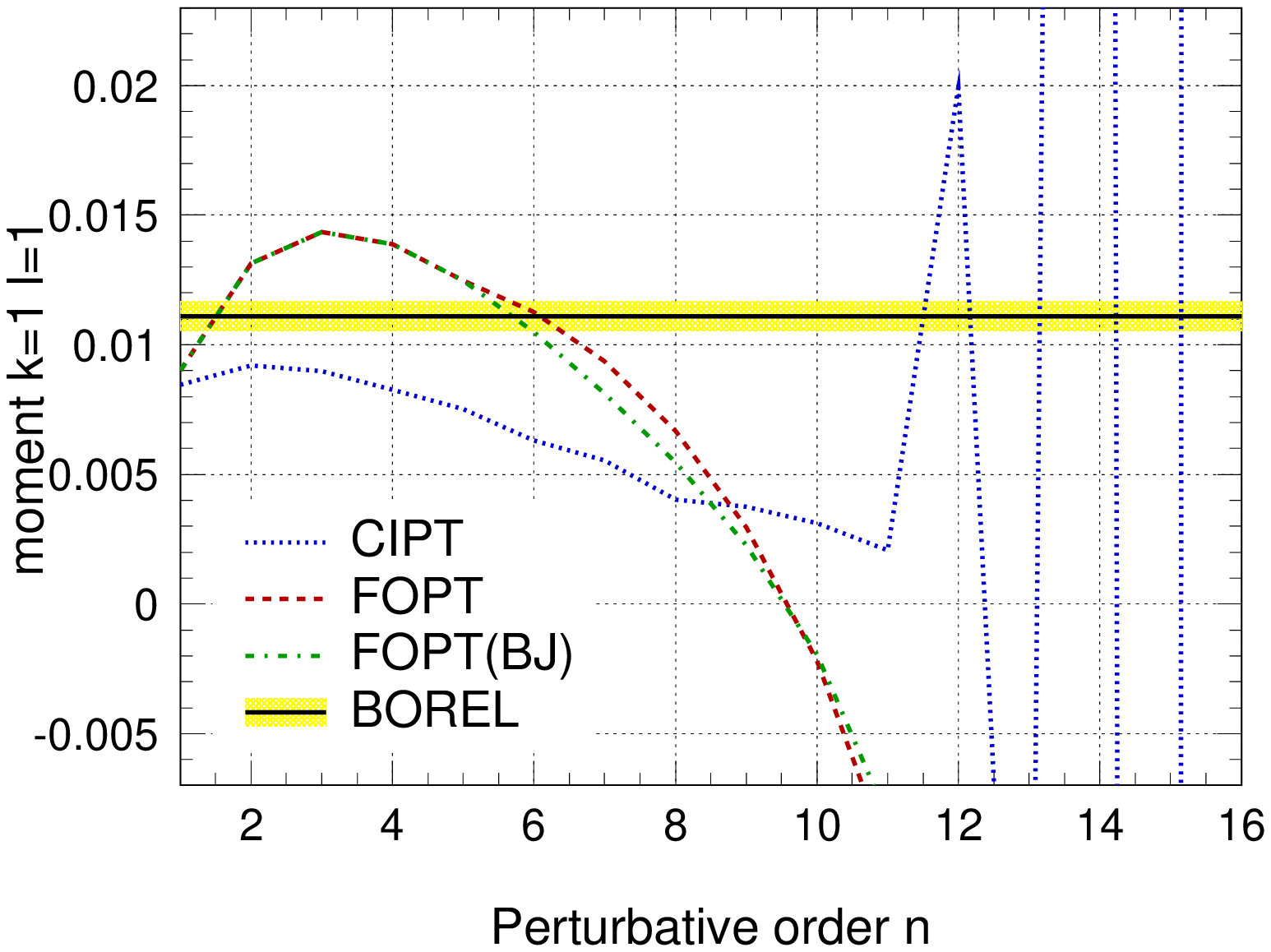}
\caption{Figures for $u_{kl}$ moments with $k=1,l=0$ (left) and $k=1,l=1$ (right).}
\label{klmoments}
\end{center}
\end{figure}

\begin{figure}[h]
\begin{center}
\includegraphics[width=8cm]{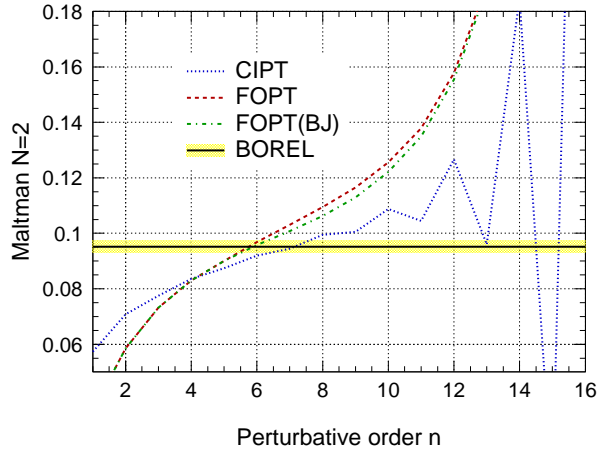}
\caption{Figure for $u_N$ moment with $N=2$.}
\label{maltman}
\end{center}
\end{figure}

One can use the same machinery to analyse further moments, for instance some higher moments used to determine the condensates from a fit to $\tau$ spectral data (fig.~\ref{klmoments})~\cite{Le Diberder:1992te,Le Diberder:1992fr,Davier:2005xq,Davier:2008sk}  and moments proposed in ref.~\cite{Maltman:2008ud} (fig.\ref{maltman}):
\begin{eqnarray*}
\label{eq:momentskl}
   R^{k=1,l}_{\tau,V+A} = &&
       \frac{6\,|V_{ud}|^2\,\Sew}{\mtau^2}\,
       \int_0^{\mtau^2} ds 
       \left(1-\frac{s}{\mtau^2}\right)^{\!\!3}\,
       \left(\frac{s}{\mtau^2}\right)^{\!\!l}\,
       \left(1+\frac{2s}{\mtau^2}\right)
       \left(v_1(s)+a_1(s)\right) \\ 
%
\label{eq:momentsM}
   R^{N}_{\tau,V+A} = &&
       \frac{6\,|V_{ud}|^2\,\Sew}{\mtau^2}\,
       \int_0^{\mtau^2} ds 
       \left[1 - \frac{N}{N-1}\frac{s}{\mtau^2} + \frac{1}{N-1}\left(\frac{s}{\mtau^2}\right)^{\!\!N}\right]\,
\end{eqnarray*}
As in the case of the $\tau$ width, we can use OPE to expand these moments. If we focus on the perturbative contribution ($D=0$), we can reexpress it as an integral over the circle using integration by part:
\begin{eqnarray}
1+\delta^{(0)}&=&-2\pi i\oint_{|s|=s_0} \frac{ds}{s} w(s) [D(s)]_{D=0}\\
   &=& 2\pi i\oint_{|s|=s_0} ds u(s) [\Pi^{(1+0)}]_{D=0}(s)=4\pi^2 \int_0^{s_0} u(s)
 [{\rm Im} \ \Pi^{(1+0)}(s)]_{D=0} \nonumber
\end{eqnarray}
where $w(s)=\int_{s_0}^s ds'\ u(s')$, so that we have
\begin{eqnarray}
u_{0l}(s)&=&-\frac{2}{s_0}\left(1-\frac{s}{s_0}\right)^2 \left(1+2\frac{s}{s_0}\right)
   \left(\frac{s}{s_0}\right)^l\\
w_{0l}(s)&=&\frac{12}{(l+1)(l+3)(l+4)}-\frac{2}{l+1}\left(\frac{s}{s_0}\right)^{l+1}
   +\frac{6}{l+3}\left(\frac{s}{s_0}\right)^{l+3}
   -\frac{4}{l+4}\left(\frac{s}{s_0}\right)^{l+4}
\end{eqnarray}
\begin{eqnarray}
u_{1l}(s)&=&-\frac{2}{s_0}\left(1-\frac{s}{s_0}\right)^3 \left(1+2\frac{s}{s_0}\right)
   \left(\frac{s}{s_0}\right)^l\\
w_{1l}(s)&=&\frac{12(3l+7)}{(l+1)(l+2)(l+3)(l+4)}-\frac{2}{l+1}\left(\frac{s}{s_0}\right)^{l+1}
   +\frac{2}{l+2}\left(\frac{s}{s_0}\right)^{l+2}\\
&&\qquad \nonumber
   +\frac{6}{l+3}\left(\frac{s}{s_0}\right)^{l+3}
   -\frac{10}{l+4}\left(\frac{s}{s_0}\right)^{l+4}
   +\frac{4}{l+5}\left(\frac{s}{s_0}\right)^{l+5}
\end{eqnarray}
\begin{eqnarray}
u_{N}(s)&=&\frac{1}{s_0}
  \left[1-\frac{N}{N-1}\frac{s}{s_0}+\frac{1}{N-1}\left(\frac{s}{s_0}\right)^N\right]\\
w_{N}(s)&=&-\frac{N}{2(N+1)}+\frac{s}{s_0}-\frac{N}{2(N-1)}\left(\frac{s}{s_0}\right)^2
   +\frac{1}{N^2-1}\left(\frac{s}{s_0}\right)^{N+1}
\end{eqnarray}
The moments $u_{N}$ were introduced in ref.~\cite{Maltman:2008ud} to suppress the higher dimensional condensates that were noted to affect the analysis of the pinched weight moments in refs.~\cite{Le Diberder:1992te,Le Diberder:1992fr,Davier:2005xq,Davier:2008sk}. These moments $u_N$  were used to extract the strong coupling constant by fitting the tau data, once the quark and gluon condensate were set to fixed values. The authors extracted the information on two different quantities (the strong coupling constant and a high-dimension condensate) by fitting
the integrals $I^N(s_0)$ obtained with the same weight $u_N$ but different radii for the contour of integration $s_0$ (between 2.3 GeV$^2$ and $m_\tau^2$). The authors claimed
an impressive agreement between the values of the strong coupling constant obtained for different $N$. 

Such an agreement  is not particularly surprising. Let us first of all notice that the points from $\tau$ data between $s^*=2.3$ GeV$^2$ and $m_\tau^2$ are correlated and have significant uncertainties, meaning that the input for the fit is essentially one integral, say $I^N(s^*)$, the integrals for other values of $s_0$ carrying very little additional information. The fit is therefore perfect, with one input and two parameters (the strong coupling constant and a high-dimensional condensate). Moreover, the output of the fit is indeed very stable as far as the strong coupling constant is concernend, since these weights can be rewritten as:
\begin{equation}
u_N(s)=\frac{1}{s_0}\left[\left(1-\frac{s}{s_0}\right)+\frac{1}{N-1}\left(\left(\frac{s}{s_0}\right)^N-\frac{s}{s_0}\right)\right]
\end{equation}
Once inserted in the integral used to computed $\delta^{(0)}$, and taking only power corrections (without logarithms) for the the OPE of $\Pi$, one can see that the first bracket provides essentially a correlation between $\alpha_s$ and the dimension-four condensates, and this correlation is identical for all the values of $N$. The second one fixes the value of the condensate of dimension $2N+2$ in terms of the dimension-4 condensate. Two sum rules for two different values of $N$ provide therefore the same correlation between $\alpha_s$ and the gluon condensate, fixed in the analysis of ref~\cite{Maltman:2008ud}.

When we compare the figures~\ref{klmoments} and \ref{maltman} for the different moments,
it is not clear whether CIPT or FOPT should be preferred in such a context. The moments tend to put a different emphasis between the contribution from $d=4$ and $d=6$ renormalons, which alter the discussion followed previously for $\delta^{(0)}$. It is quite interesting to notice that the agreement between FOPT and the Borel sum is not automatic, and depends on the structure of the kernel considered. In order to quantify this, one can consider the difference between the contribution of the IR poles in FOPT/CIPT and their Borel resummed values, as shown in Table~\ref{table:vars} for $u_{00},u_{10},u_{11}$ and $u_2$. The agreement with the Borel resummed version for $n$ around 7 is better for CIPT in the case of $u_2$,  equally bad for FOPT and CIPT in the case of $u_{kl}$ for $k=1,l=0$, and better for FOPT in the case of $k=1,l=1$.

\begin{table*}[h]
  \caption[.]{\label{table:vars}
              Results for the discrepancy between contributions for IR poles in FOPT/CIPT computations and the Borel sum for various weights}
\begin{center}
\setlength{\tabcolsep}{0.0pc}
\begin{tabular*}{\textwidth}{@{\extracolsep{\fill}}ccrrrrrrrrr} 
\hline\noalign{\smallskip}
& IR   &   $4$   &   $5$   &   $6$   &   $7$   &   $8$   &   $9$   &   $10$\\
\noalign{\smallskip}\hline\noalign{\smallskip}
$k,l=0,0$ &
2 & -0.028952 & -0.013646 & -0.000064 & 0.007366 &
 0.007794 & 0.003553 & -0.001291\\
FOPT &
3 & -0.003477 & -0.001710 &  0.000195 & 0.001334 &
 0.001405 & 0.000674 & -0.000239\\
\hline
$k,l=0,0$ &
2 & -0.012736 & -0.011107 & -0.010368 &-0.010359 & 
-0.011042 & -0.012507 & -0.014986\\
CIPT  &
3 & -0.000084 & -0.000028 & -0.000004 &
      -0.000004 & -0.000017 & -0.000040 &
      -0.000067 \\
\noalign{\smallskip}\hline\hline\noalign{\smallskip}
$k,l=1,0$ & 2 & -0.013614 & -0.009428 & -0.003829 & 0.001039 &
0.004036 & 0.005141 & 0.005348\\
FOPT & 3 & -0.001338 & -0.001083 & -0.000471 &
0.000110 & 0.000423 & 0.000406 & 0.000167\\
$k,l=1,0$ &
2 & -0.006784&  -0.005691 & -0.005018 &
-0.004723 & -0.004830 & -0.005442 & -0.006781\\
CIPT & 3 & -0.000165 & -0.000127 & -0.000106 &
-0.000098 & -0.000100 & -0.000109 &
-0.000124\\
\noalign{\smallskip}\hline\hline\noalign{\smallskip}
$k,l=1,1$ &
2 & -0.002037&  0.000328 & 0.001658 & 0.001569 &
0.000149 & -0.002133 &  -0.004686\\
FOPT & 3 & -0.000389 & 0.000017 & 0.000317 &
0.000423 & 0.000337 & 0.000131 & -0.000076\\
$k,l=1,1$ &
2 & -0.000323 &  -0.000647 & -0.001017& 
-0.0014017 & -0.001768&  -0.002060& -0.002160\\
CIPT & 3   & 0.000131 & 0.000120 & 0.000108 &
0.000098 & 0.000090 & 0.000086 & 0.000085\\
\noalign{\smallskip}\hline\hline\noalign{\smallskip}
$N=2$ &
2 & -0.009792 &  -0.006531 & -0.002007 & 0.002078 & 
0.004959 & 0.006847 & 0.008774\\
FOPT & 3 & -0.000934 & -0.000814 & -0.000391& 
0.000014 & 0.000222 & 0.000206 &  0.000066\\
$N=2$ &
2 & -0.004806 & -0.003168 &  -0.001764 &
-0.000503 & 0.000692 & 0.001890 & 0.003152\\
CIPT &
3 & -0.000243 & -0.000186 & -0.000141 & 
-0.000109 & -0.000085 & -0.000067 &
-0.000053\\
\noalign{\smallskip}\hline
\end{tabular*}
  \end{center}
\end{table*}

\section{Conclusion}

In this paper, we have investigated several aspects of renormalon models for the Adler function, used recently to compare fixed-order and contour-improved perturbation theories (FOPT and CIPT) to treat the contour integral for the theoretical estimate of the $\tau$ width~\cite{Beneke:2008ad,Caprini:2009vf} . Indeed the difference between the two treatments induce a significant systematics on the extraction of the
strong coupling constant at the $\tau$ mass. The particular renormalon ansatz for the Adler function in ref.~\cite{Beneke:2008ad} suggested that FOPT was to be preferred with respect to CIPT, since it converges to the value of the Borel sum (taken as the true value of the integral).

During our study, we have noticed the following points:
\begin{itemize}
\item Once several infrared poles are included in the model, one needs to define how one combines the uncertainties estimated from the Borel integral. Depending on the treatment of the RGE for $\alpha_s$ at higher orders (for which the $\beta$ function is not known), the FOPT prescription can yield noticeable differences at intermediate orders, in agreement with more conservative estimates of the uncertainty.
\item  Ref.~\cite{Beneke:2008ad} sets to zero the contributions from unknown terms in the Wilson coefficients involved in the OPE of the Adler function . Moreover, the perturbative expansion of the ansatz is obtained through
an expansion in powers of $1/n$ (where $n$ is the order of perturbation theory) where only the first order are kept.
This (truncated) perturbative expression is used to determine the residues of the poles from the first order of the perturbative expansion of $D$. We noticed that the neglected contributions are not particularly small at the orders of perturbation theory used to determine these residues, which can affected by significant uncertainties.
\item We extended the renormalon models by taking into account such potentially large contributions in particular for $n=3$, and we investigated some cases where FOPT or CIPT are in better agreement with the values obtained from the Borel integral. We discussed two different definitions of the dominance of a pole, in order to determine which cases of these extended models could be considered as acceptable because of the dominance of the first infrared pole. 
\item We examined the issue of the anomalous dimension of the second infrared pole, which actually corresponds to five operators of different dimensions, and observed rather different behaviours of the perturbative series.
\item We discussed other weights, noticing that the better agreement of FOPT with the value of the Borel integral is not a universal feature, and depends on which part of the contour integral is suppressed or enhanced by the weight.
\end{itemize}

Renormalon models provide very attractive features to discuss qualitative aspects of higher-order perturbation theory. However, in the present discussion, we want to compare small differences between two treatments of perturbation theory, requiring a quantitative model of the higher orders of the Adler function. Any given ansatz based on renormalon calculus involve a large number of unknown coefficients both for the singular terms (residues of the renormalon "poles") and the non-singular terms (polynomial contribution). Only a limited number of these coefficients can be fixed through the first few known terms of the perturbative expansion of the Adler function -- the other ones being generally set to zero. It is not clear that the simplified description of the renormalon singularities by an ansatz, assumed to be valid at high orders, is sufficient at such low orders. 

The particular ansatz chosen in ref.~\cite{Beneke:2008ad} does not exhaust the potentialities of model building provided by renormalon calculus, and we have described a few extensions leading to rather varied conclusions concerning the comparison of FOPT versus CIPT. Our study shows that this particular ansatz cannot be taken as part of a reference test to determine
whether FOPT, CIPT or yet another method should be adopted to extract $\alpha_s$ from hadronic $\tau$ decays. Moreover, significant systematics (as large as the difference between the standard CIPT and FOPT results) ought to be added to the results based on such an ansatz, since it is only one among many different renormalon models for perturbative expansions at high orders. 

The previous discussion is essentially based on the fact that we assume the Borel sum eq.~(\ref{eq:borelsum}) to provide the "true" value of the asymptotic perturbative series for the Adler function. In particular, its value is used to determine whether FOPT or CIPT should be preferred. Let us mention that the theoretical estimation of the $\tau$ decay is rather particular in this respect, since the low value of the $\tau$ mass compared to hadronic scales requires one to compute an integral over a contour in the complex energy plane. The assumption that the Borel sum yields the true value of the Adler function should hold not only for real positive values of the coupling constant, but also for values of $\alpha_s(s)$ where $s$ is complex. The theory of asymptotic expansions~\cite{Dingle} indicates that there are functions for which this continuation of asymptotic series is not simple and one might encounter discontinuities when one crosses frontiers in the complex plane (Stokes lines). It would be interesting to determine whether such a situation could occur in renormalons models, and what their impact could be in the issues discussed here.

\acknowledgments

We would like to thank M. Beneke, M. Davier, S. Friot, A. H\"ocker, M. Jamin, T. Pich and Z. Zhang for stimulating discussions. This work is supported in part by the EU Contract No. MRTN-CT-2006-035482, "FLAVIAnet".

\end{document}